\documentclass[journal]{IEEEtran}
\usepackage{amsmath, amssymb}
\usepackage{algorithm, algpseudocode}
\usepackage{array}
\usepackage{graphicx}
\usepackage{bm}
\usepackage{cite}
\usepackage{booktabs}
\usepackage{multirow}
\usepackage{placeins}

\begin{document}
\title{A Message-Passing Perspective on Ptychographic Phase Retrieval}

\author{Hajime Ueda, Shun Katakami, Masato Okada
\thanks{H.Ueda, S. Katakami, and  M. Okada are with Graduate School of Frontier Sciences, The University of Tokyo, Japan}
\thanks{H.Ueda is supported by JST BOOST, Japan Grant Number JPMJBS2418. This work is partly supported by JSPS KAKENHI Grant Number 23H00486, 23K16959, and Digital Transformation Initiative for Green Energy Materials (DX-GEM).}}% <-this % stops a space

\maketitle
\begin{abstract}
We introduce a probabilistic approach to ptychographic reconstruction in computational imaging.
Ptychography is an imaging method where the complex amplitude of an object is estimated from a sequence of diffraction measurements.
We formulate this reconstruction as a Bayesian inverse problem and derive an inference algorithm, termed "Ptycho-EP," based on belief propagation and Vector Approximate Message Passing from information theory.
Prior knowledge about the unknown object can be integrated into the probabilistic model, and the Bayesian framework inherently provides uncertainty quantification of the reconstruction.
Numerical experiments demonstrate that, when the probe's illumination function is known, our algorithm accurately retrieves the object image at a sampling ratio approaching the information theoretic limit.
In scenarios where the illumination function is unknown, both the object and the probe can be jointly reconstructed via an Expectation-Maximization algorithm.
We evaluate the performance of our algorithm against conventional methods, highlighting its superior convergence speed.\end{abstract}

\section{Introduction}
\subsection{Ptychography and Belief Propagation}
\IEEEPARstart{P}{tychography} is an advanced imaging technique that combines diffraction measurements and computational algorithm to reconstruct the complex amplitude of an object.
Since the development of iterative solvers and the experimental validation in X-ray microscopy\cite{PRL_PIE,APL_PIE,ePIE,DifferenceMap}, it has evolved into a foundation for modern computational imaging methods, such as 3D ptychography~\cite{3PIE,PtychoTomo} and Fourier ptychography~\cite{FourierPtycho_Nature,FourierPtycho_concept}.
Today, this method is applied in a wide range of scientific fields, including the quantitative imaging of atomic structures, nano materials, and biological specimens~\cite{miao_computational_2025}.

Ptychography is regarded as a variant of Coherent Diffraction Imaging (CDI) that employs multiple measurements, where each diffraction dataset, indexed by $j = 1,2,..., J$, is acquired by illuminating distinct regions of a two-dimensional object, as depicted in Fig. 1.
The reconstruction algorithm takes as input the $J$ diffraction datasets along with their corresponding scan coordinates, and the aim is to estimate the unknown object image from these data.
Depending on the experimental conditions, the illumination function of the coherent wave \cite{ePIE,DifferenceMap} and errors in scan positions\cite{PositionCorrection2012,PositionCorrection2013} are jointly recovered with the object.
Furthermore, the forward model of ptychography has been extended for three-dimensional objects \cite{3PIE} and illumination with partially coherent beam\cite{MixedStates2013}.

\begin{figure}[htb]
  \begin{minipage}[b]{1.0\linewidth}
    \centering
    \centerline{\includegraphics[width=8.5cm]{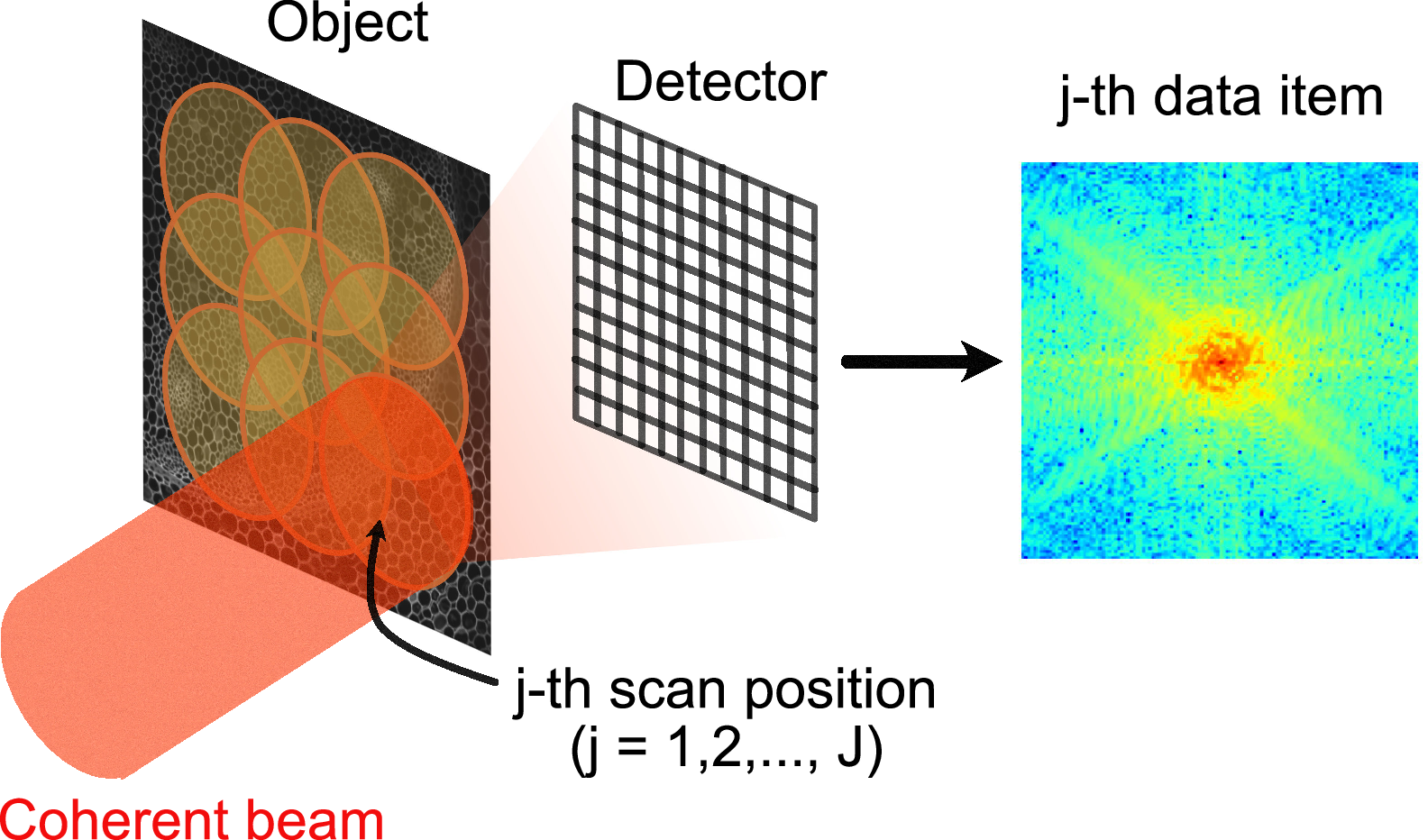}}
    \caption{Illustration of 2-D ptychography experiment.}\medskip
  \end{minipage}
\end{figure}

To illustrate the problem setting of ptychographic reconstruction, we examine the case where the number of scans $J = 3$.
Each of the three illuminated regions is depicted as a disk in Fig.1.(a), while the complex vectors $\bm{O}_1, \bm{O}_{12}, \cdots$ correspond to the object images within the overlapping regions.
For instance, the data item 1 is expressed as a nonlinear function of the vector set $(\bm{O}_1, \bm{O}_{12}, \bm{O}_{13}, \bm{O}_{123})$.
In Fig.1.(b), white circles and black squares denote variables and measurements, and each variable node is linked to corresponding measurement nodes.
This forms the factor graph that represents the inference problem in ptychography.

As an instance of similar problem, we show a graphical model of Low Density Parity Check (LDPC) code\cite{LDPC} from coding theory, in Fig. 2. (c).
In this case, the variables $x_1, x_2, \cdots$ are binary, and the data items in Fig. 2. (b) are replaced by constraints $C_1, C_2, C_3$.
For example, since the node $C_1$ is linked to the nodes $x_2, x_3, x_4$ and $x_5$, it imposes the constraint $x_2 + x_3 + x_4 + x_5 = 0$.
The objective is to recover the binary sequence $\bm{x} = (x_1,x_2,..., x_6)$ that satisfies these constraints, given a noise-corrupted observation of $\bm{x}$.
It is well established that LDPC codes achieve rates approaching the Shannon limit and can be efficiently decoded through iterative algorithms based on Belief Propagation (BP)\cite{BP}.
In essence, our approach is to use BP to solve ptychographic reconstruction.

\begin{figure*}[t]
  \begin{minipage}[b]{1.0\linewidth}
    \centering
    \centerline{\includegraphics[width=7cm, angle = 90]{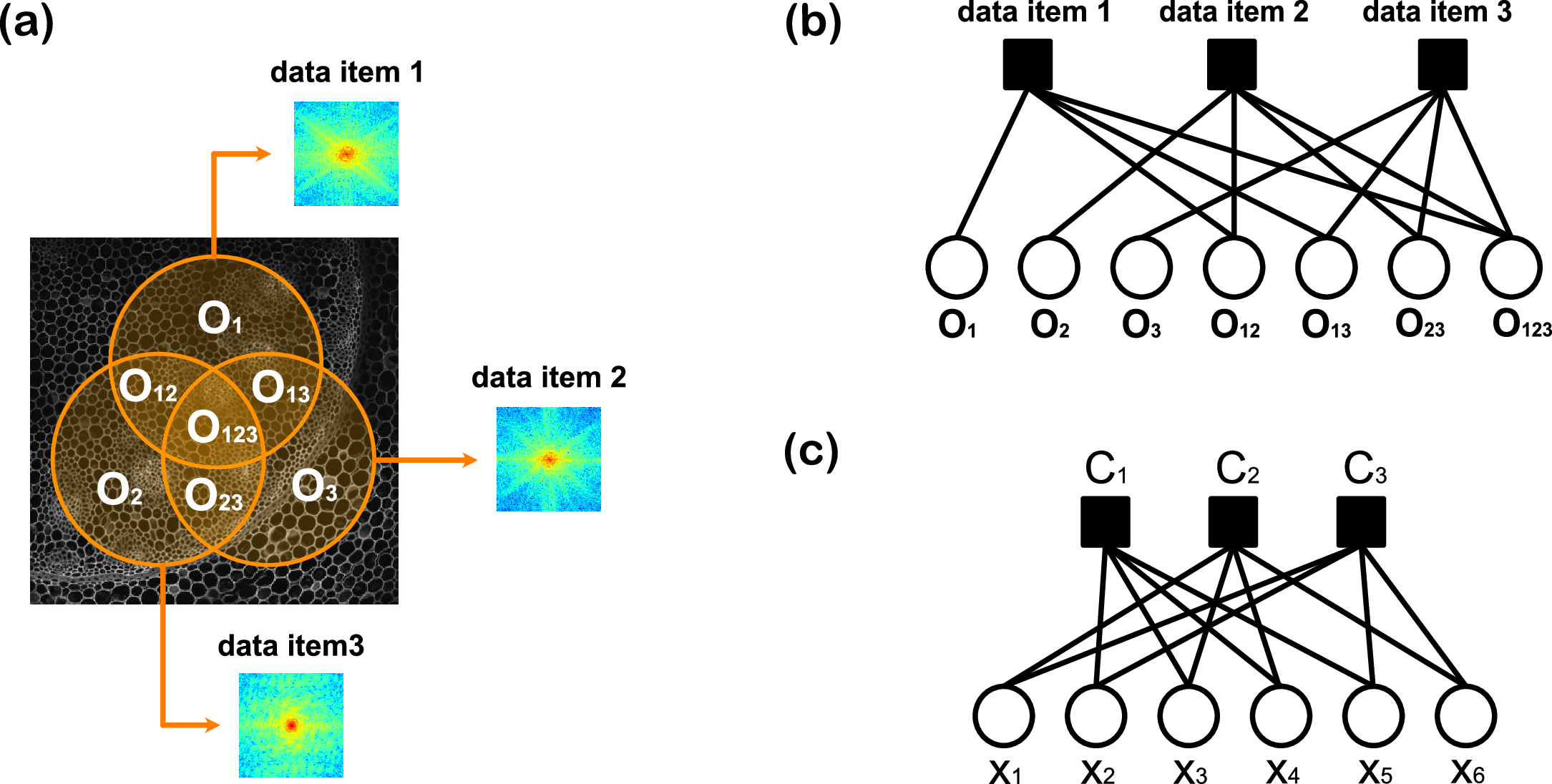}}
    \caption{(a) Ptychography with $J = 3$ scan positions. 
    (b) The graphical representation of ptychographic reconstruction, where each circle denotes a variable node, and the black square signifies a factor node.
     (c) The graphical model of [2,4] LDPC code. The factor nodes $C_1, C_2, C_3$ impose constraints on the binary variables $x_1, x_2, \cdots, x_6$. }\medskip
  \end{minipage}
\end{figure*}

Within the framework of BP, probabilistic distributions, referred to as "messages", are transmitted by both variable (white circle) and factor (black square) nodes, along the edges of the factor graph.
Those messages are iteratively updated following the principles of sum-product algorithm\cite{sumproduct} until convergence.
BP provides an estimate of marginal distribution for each variable, expressed as the normalized product of the incoming messages to the variable node.
This computation yields exact marginals when the graph has no loops\cite{mezard_information_2009}.

\subsection{Phase Retrieval}
A crucial aspect of ptychographic reconstruction is that, since the detector cannot observe the phase of the coherent wave, one has to solve the phase problem\cite{Fienup1982}.
Formally, a phase problem is stated as the estimation of a complex vector $\bm{O} \in \mathbb{C}^N$ from the intensity-only measurement of $\bm{\Phi} \in \mathbb{C}^M$, given by
\begin{align}
  \bm{\Phi} = \mathsf{A}\bm{O}
\end{align}
We refer to the matrix $\mathsf{A} \in \mathbb{C}^{M \times N}$ as the measurement matrix. The observation vector $\bm{I} \in \mathbb{R}^M$ is the noisy version of $|\bm{\Phi}|^2$, where $|\cdot|$ is the component-wise absolute value.
Solvers for phase problem is called phase retrieval algorithms, and a plethora of methods have been proposed, based on Wirtinger Flow\cite{WirtingerFlow}, alternating projection\cite{Fienup1982,Gerchberg1972APA}, convex relaxation\cite{PRviaMC,PhaseLift,phasemax}, and spectral initialization\cite{Luo2018,Mondelli2017}.
In our approach, we incorporate BP with a phase retrieval algorithm based on Vector Approximate Message Passing (VAMP) \cite{VAMP,GVAMP} , an efficient inference algorithm for generalized linear models.

To fully appreciate the state-of-the-art performance of VAMP, we briefly review on the hardness of phase retrieval as an inverse problem.
Since phase retrieval algorithms estimate an $N$- dimensional complex vector from an $M$- dimensional real vector, the sampling ratio $\alpha = M/N$ larger than $\alpha_{\text{stat}} = 2$ should be sufficient to uniquely determine the solution.
Unfortunately, this does not mean that polynomial-time algorithms can find the solution always when $\alpha > \alpha_{\text{stat}}$.
In the setting where $\mathsf{A}$ is a large random matrix, the algorithmic threshold $\alpha_{\text{alg}} (> \alpha_{\text{stat}})$ is identified for phase problem, beyond which VAMP can accurately estimate $\bm{O}$ with probability one in the $N \rightarrow \infty$ limit~\cite{Maillard2020}.
Notably, no known polynomial-time algorithm is capable of solving phase problem when $\alpha < \alpha_{\text{alg}}$ in this setting, motivating the extension of VAMP to ptychographic reconstruction.
This discrepancy between the statistical threshold $\alpha_{\text{stat}}$ and the algorithmic threshold $\alpha_{\text{alg}}$ is common across a wide range of inverse problems~\cite{Bandeira2018NotesOC}.

The algorithmic threshold $\alpha_{\text{alg}}$ is primarily determined by the character of the measurement matrix $\mathsf{A}$.
For example, if each entry of $\mathsf{A}$ is independently drawn from complex Gaussian distribution $\mathcal{CN}(0, 1/N)$, the algorithmic threshold is $\alpha_{\text{alg}} \simeq 2.027$, whereas the value for a randomly generated orthogonal matrix (i.e. Haar distributed matrix) is $\alpha_{\text{alg}} \simeq 2.265$\cite{Maillard2020}.
While these results premise on the assumption that $\mathsf{A}$ is drawn from a right-invariant random matrix ensemble~\cite{VAMP}, measurement matrices encountered in real-world applications are far from this model.
Indeed, VAMP does not exhibit practical performance when applied to deterministic matrices such as discrete Fourier transform (DFT).
Remarkably, there are some "randomization" techniques that make deterministic matrices amenable to random-matrix-based algorithms such as VAMP.
A typical example of this strategy is the Coded Diffraction Pattern~\cite{CodedDiffractionPattern}, wherein the measurement matrix is given as
\begin{align}
  \mathsf{A} = 
  \begin{pmatrix}
    \mathsf{F} \ \mathsf{Diag}(\bm{P}_1) \\
    \vdots \\
    \mathsf{F} \ \mathsf{Diag}(\bm{P}_J) \\
  \end{pmatrix}
\end{align}
Here, $\mathsf{F} \in \mathbb{C}^{N \times N}$ represents the 2-D DFT matrix and each entry of $\bm{P}_1,...,\bm{P}_J \in \mathbb{C}^N$ is independently drawn from the uniform distribution on the unit circle in $\mathbb{C}$.
This measurement matrix can be seen as a model of Coherent Diffraction Imaging with randomized illumination, which is a technique actively utilized in computational imaging community~\cite{Randomillumination,Horisaki:16,random_phase_ptychography}.
In this work, we numerically demonstrate that ptychographic reconstruction is possible under the sampling ratio close to the algorithmic limits for random matrices, which indicates that ptychography can be seen as another technique to randomize DFT matrix.

\begin{figure*}[]
  \begin{minipage}[b]{1.0\linewidth}
    \centering
    \centerline{\includegraphics[width=18cm]{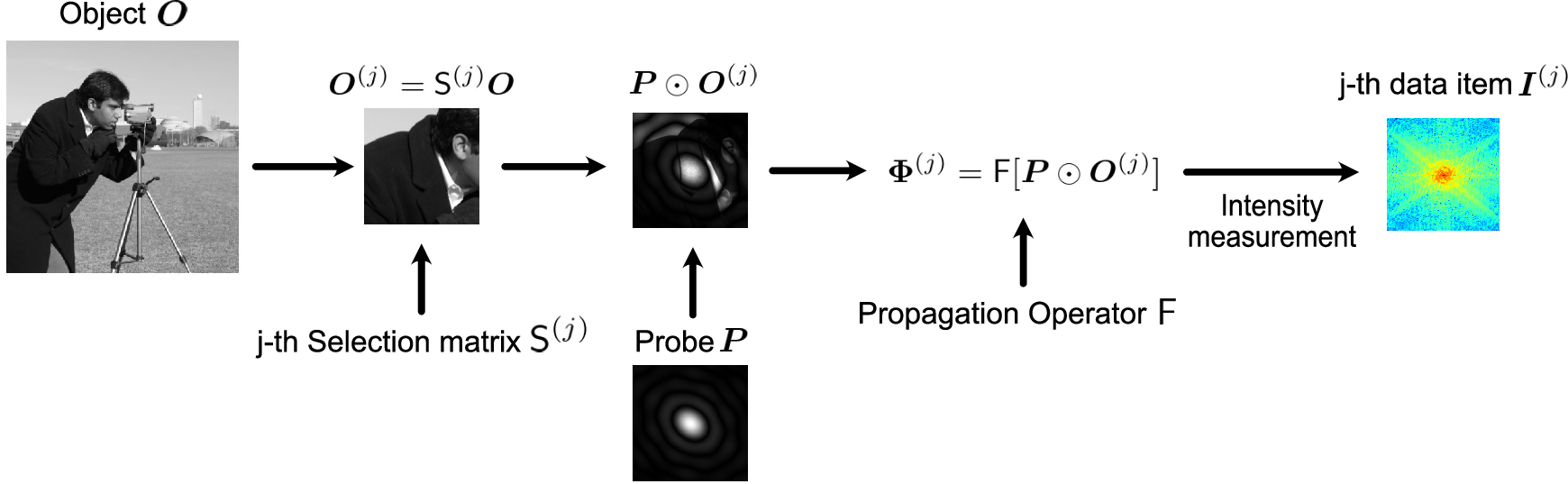}}
    \caption{The forward model of the diffraction datasets acquired in 2-D ptychography. $j \in \{1,2,...,J\}$ refers to the index for each diffraction measurement.}\medskip
  \end{minipage}
\end{figure*}

\subsection{Overview of the Paper}
The objective of this paper is to introduce a novel approach to ptychographic reconstruction, grounded in the principles of belief propagation.
In Section II, we formulate the Bayesian inverse problem that we consider throughout this study.
Section III describes the derivation of our algorithm, Ptycho-EP, based on the graphical model established in section II.
We also address the problem of ptychographic reconstruction with unknown probe, for which an Expectation-Maximization (EM) algorithm is employed to jointly estimate both the object and the probe.
Section IV presents computational illustration of our algorithm, both in known and unknown probe case.
When the probe is known apriori, we show that Ptycho-EP can recover an unknown object under a sampling ratio approaching the information-theoretic limit.
Furthermore, we demonstrate that data redundancy can be reduced by incorporating a sparsity prior.
The performance of Ptycho-EP with probe-retrieval is examined in a comparison with conventional methods, varying the overlap ratio between illuminated spots. Our numerical experiments demonstrates the superior convergence speed of Ptycho-EP.
Concluding remarks will be made in section V, and detailed explanations on algorithm modules utilized in Ptycho-EP are provided in appendices.

We summarize the notation employed in this paper.
Vectors are represented using bold uppercase ($\bm{O}, \bm{P}, \bm{\Psi}, \bm{\Gamma}\cdots$) in accordance with conventions in ptychography literature.
Given a vector $\bm{A} \in \mathbb{C}^N$ , $\bm{A}^n$ ($n \in \mathbb{Z}$) and $|\bm{A}|$ represent component-wise absolute value and exponentiation, respectively. 
We denote the $l_2$ norm of $\bm{A}$ by $\| \bm{A} \|_2$.
For $\bm{A} = (A_1,\cdots,A_N)$, $\text{tr}(\bm{A})$ is defined by $\text{tr}(\bm{A}) = A_1+\cdots+A_N$.
For two vectors $\bm{A}$ and $\bm{B}$, the operations $\bm{A} \odot \bm{B}$ and $\frac{\bm{A}}{\bm{B}}$ refer to Hadamard multiplication and division, respectively.
Matrices are represented using serif uppercase letters  ($\mathsf{A}, \mathsf{F},\cdots$).
$\bm{0}$ and $\bm{1}$ refer to the vector with all zeros and all ones respectively, and $\mathsf{I}_N$ denotes the identity matrix.
The notation $\mathsf{Diag}(\bm{A})$ refers to the diagonal matrix whose diagonal elements are given by the vector $\bm{A}$, while $\text{diag}(\mathsf{C})$ represents the diagonal matrix obtained by zeroing all of the off-diagonal components of matrix $\mathsf{C}$.
For vectors and matrices, superscripts $^T$, $^H$, and $^*$ denote transpose, Hermite transpose, and component-wise complex conjugate, respectively.
For a complex vector $\bm{M} \in \mathbb{C}^N$ and a positive semi-definite matrix $\mathsf{\Sigma} \in \mathbb{C}^{N \times N}$, the Gaussian distribution $\mathcal{CN}(\bm{M}, \mathsf{\Sigma})$ is defined as a measure on $\mathbb{C}^N$ with density
\begin{align}
  \mathcal{CN}(\bm{X}; \bm{M}, \mathsf{\Sigma}) \propto \exp \left(-(\bm{X} - \bm{M})^H \mathsf{\Sigma}^{-1} (\bm{X} - \bm{M}) \right).
\end{align}

\section{Problem setting}
Figure 3 illustrates the forward model of 2-D ptychography.
The object $\bm{O} \in \mathbb{C}^N$ is a $\sqrt{N} \times \sqrt{N}$-pixel image with complex values.
$J \in \mathbb{Z}_+$ denotes the number of diffraction measurements, and the selection matrix $\mathsf{S}^{(j)} \in \{0,1\}^{M \times N}$ is given for each index $j \in \{1,2,..., J\}$.
This matrix selects a $\sqrt{M} \times \sqrt{M}$-pixel region in the object image $\bm{O}$, which is the area illuminated by the probe at the $j$-th measurement.
We denote the selected sub-image by $\bm{O}^{(j)} \triangleq \mathsf{S}^{(j)} \bm{O}$.
The probe is represented by a $\sqrt{M} \times \sqrt{M}$-pixel image $\bm{P} \in \mathbb{C}^M$.
When the $j$-th position of the object is illuminated by the probe, the coherent wave that exits the 2-D object is modeled by $\bm{P} \odot \bm{O}^{(j)}$.
The complex amplitude of the wave that arrives at the detector is given as
\begin{align}
  \bm{\Phi}^{(j)} = \mathsf{F}[\bm{P} \odot \bm{O}^{(j)}]
\end{align}
where the propagation operator $\mathsf{F} \in \mathbb{C}^{M \times M}$ is a unitary matrix that is determined by the experimental setting.
For example, the operator for Fraunhofer diffraction is 2-D DFT, and the angular spectrum operator is employed for Fresnel diffraction~\cite{FourierOptics}.

The observed data $\bm{I}^{(j)} \in \mathbb{R}^M$ is the noise-corrupted version of $|\bm{\Phi}^{(j)}|^2$.
The distribution of $\bm{I}^{(j)}$ given $\bm{\Phi}^{(j)}$ is denoted by $p_{\text{out}}(\bm{I}^{(j)} | |\bm{\Phi}^{(j)}|^2)$.
If the noise is modeled by the counting statistics, $p_{\text{out}}$ is given as
\begin{align}
  p_{\text{out}}(\bm{I}^{(j)} | |\bm{\Phi}^{(j)}|^2) = Po(\bm{I}^{(j)} ; |\bm{\Phi}^{(j)}|^2)
\end{align}
where $Po(x ; \lambda) \propto \lambda^x / x!$ is the probability mass function of Poisson distribution.
A common approximation to Eq. (5) is 
\begin{align}
  p_{\text{out}}(\bm{I}^{(j)} | |\bm{\Phi}^{(j)}|^2) &= \mathcal{N}(\sqrt{\bm{I}^{(j)}} ; |\bm{\Phi}^{(j)}|, \sigma^2 I_M)\\\nonumber
 &\propto \exp \left(-\frac{1}{2\sigma^2} \|\sqrt{\bm{I}^{(j)}} - |\bm{\Phi}^{(j)}|\|_2^2 \right)
\end{align}
with $\sigma = 1/2$. Practical phase retrieval algorithms such as Fienup's algorithm~\cite{Fienup1982} can be viewed as the minimization of the error $-\log p_{\text{out}}(\bm{I}^{(j)} | |\bm{\Phi}^{(j)}|^2)$ defined by Eq. (6).

To establish a Bayesian framework, we assume a probabilistic structure of the unknown object $\bm{O}$.
One may use, for instance, Gaussian prior $p_{\text{in}}(\bm{O}) = \mathcal{CN}(\bm{0}, I_N)$ when no prior information is accessible, as in the previous work on Bayesian phase retrieval by variational approach~\cite{PRVBEM}.
In the case where $\bm{O}$ is assumed to be sparse, a simple choice of prior is the Bernoulli-Gaussian (BG) model
\begin{align}
  p_{\text{in}}(\bm{O}) = \prod_{n = 1}^N \{ \rho\  \mathcal{CN}(O_n ; 0,1) + (1-\rho) \delta (O_n)\}
\end{align}
wherein $\rho \in (0,1)$ is the sparsity rate. This prior is used in ~\cite{PRGAMP} to achieve phase recovery from limited size of diffraction data.
As will be explained later, those inference methods exploit prior knowledge only through the denoising function 
\begin{align}
  \bm{g}_{\text{in}}(\widetilde{\bm{O}}, \mathsf{\Sigma}) \triangleq \frac{\int \bm{O} P_{\text{in}}(\bm{O}) \mathcal{CN}(\bm{O} ; \widetilde{\bm{O}}, \mathsf{\Sigma} ) d\bm{O}}{\int  P_{\text{in}}(\bm{O}) \mathcal{CN}(\bm{O} ; \widetilde{\bm{O}}, \mathsf{\Sigma} ) d\bm{O}}
\end{align}
wherein $\widetilde{\bm{O}} = \bm{O} + \bm{\epsilon}$ is the observation of $\bm{O}$ corrupted by noise $\bm{\epsilon} \sim \mathcal{CN}(\bm{0}, \mathsf{\Sigma})$, and $g_{\text{in}}(\bm{R}, \mathsf{\Sigma})$ is the MMSE estimator of $\bm{O}$.
By replacing Eq.(8) with common denoising algorithms such as deep-net denoisers, the features of natural images can be utilized within phase retrieval~\cite{deepEC}.
In this paper, we use Gaussian and BG priors for the demonstration of our Bayesian approach.

In summary, we consider $J+1$ random variables $\bm{O}, \bm{\Psi}^{(1)}, \bm{\Psi}^{(2)}, \cdots , \bm{\Psi}^{(J)}$ distributed according to 
\begin{equation}
  \begin{split}
    &p(\bm{O}, \bm{\Psi}^{(1)}, \cdots, \bm{\Psi^{(J)}} | \bm{I}^{(1)},\cdots,\bm{I}^{(J)}) 
    \propto p_{\text{in}}(\bm{O}) \times \\
    &\prod_{j=1}^{J}\delta(\bm{\Psi}^{(j)} - \mathsf{F}[\bm{P} \odot \bm{O}^{(j)}]) 
    \times \prod_{j=1}^J p_{\text{out}}(\bm{I}^{(j)} | |\bm{\Psi}^{(j)}|^2)
  \end{split}
  \end{equation}
The whole probabilistic model is represented by a factor graph with $J$ branches, shown in Fig. 4.
Note that, while $\bm{O}_1, \bm{O}_{12}, \bm{O}_{23},...$ in Fig.2.(b) are drawn as separate vectors, those vectors are integrated into single vector $\bm{O}$ in Fig.4 so that the graph has no loops.

\begin{figure}[htb]
  \begin{minipage}[b]{1.0\linewidth}
    \centering
    \centerline{\includegraphics[width=9cm]{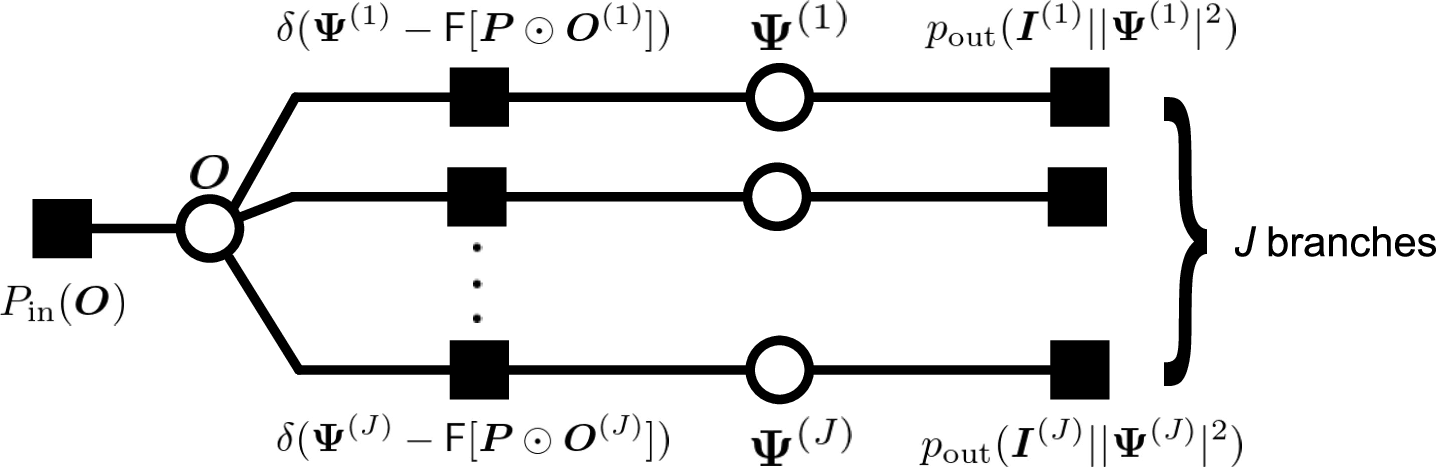}}
    \caption{The factor graph of the bayesian framework for ptychography.}\medskip
  \end{minipage}
\end{figure}

We now describe the problem we will be addressing.
The simplest form of \textit{Ptychographic phase retrieval} is stated as follows: 
\begin{align}
  \text{Estimate } \bm{O} \text{ given } \bm{P}, (\mathsf{S}^{(j)})_{j = 1}^J, (\bm{I}^{(j)})_{j=1}^J \text{, and } \mathsf{F}
\end{align}
This problem can be viewed as a special case of phase retrieval by defining the measurement matrix as
\begin{align}
  \mathsf{A}_{\text{ptycho}} = 
  \begin{pmatrix}
    \mathsf{F} \mathsf{S}^{(1)} \mathsf{Diag}(\bm{P}) \\
    \vdots \\
    \mathsf{F} \mathsf{S}^{(J)} \mathsf{Diag}(\bm{P}) 
  \end{pmatrix}.
\end{align}
The most classic approach to this problem is the PIE algorithm~\cite{APL_PIE}, where the vectors $\bm{\Psi}^{(1)},\cdots,\bm{\Psi}^{(J)}$ are sequentially projected to the set $\{\bm{\Psi}^{(j)}\in \mathbb{C}^M | |\bm{\Psi}^{(j)}| = \sqrt{\bm{I}^{(j)}}  \}$ and $\bm{O}$ is updated using $\bm{\Psi}^{(j)}$ vectors.
There are several variants of the update rule of $\bm{O}$, namely ePIE~\cite{ePIE} and rPIR ~\cite{rPIE}.

Interestingly, phase retrieval algorithms based on random matrix models are known to work well with the measurement matrix $\mathsf{A}_{\text{ptycho}}$.
For example, Wirtiger Flow algorithm~\cite{WirtingerFlow}, whose exact phase retrieval is proved for random measurement matrix, is applied to Fourier ptychography in ~\cite{FourierPtychoWF}, exhibiting state-of-the-art reconstruction quality without using informative priors.
Another example is Spectral Initialization~\cite{Mondelli2017,Luo2018}, which provides an estimate of $\bm{O}$ that is correlated with the true object image.
Although the performance of SI is guaranteed for a certain random matrix ensembles, it has been demonstrated that SI gives good initial value for iterative solvers~\cite{Valzania2021}.
These empirical results led us to the use of VAMP for ptychographic reconstruction.

In practice, obtaining an accurate knowledge of $\bm{P}$ before reconstruction is time-consuming or sometimes impossible.
Thus, the following \textit{blind ptychographic phase retrieval} needs to be addressed:
\begin{align}
  \text{Estimate } \bm{O} \text{ and } \bm{P} \text{ given } (\mathsf{S}^{(j)})_{j = 1}^J, (\bm{I}^{(j)})_{j=1}^J \text{, and } \mathsf{F}
\end{align}
In this case, we are to solve a phase problem with uncertainty in the measurement matrix. 
Two of the most popular methods are PIE algorithms with probe update~\cite{ePIE,rPIE} and the Difference Map algorithm~\cite{DifferenceMap}.
From a broader view, blind ptychographic phase retrieval is an instance of bilinear recovery, since $\bm{\Psi}^{(j)}$ is linear with respect to both $\bm{O}$ and $\bm{P}$.
Inverse problems with such a structure frequently appears in machine learning and signal processing, including matrix factorization~\cite{DictionaryLearning,NMF} and blind de-convolution~\cite{BlindDeconvolution}.
In the next section, we will state our approaches to the two problems defined in (10) and (12).

\section{Ptycho-EP}
The methodology underpinning our approach is the G-VAMP~\cite{GVAMP,Maillard2020} algorithm, a unified framework designed for Bayesian inference in generalized linear models.
While Bayesian approaches to phase retrieval have been explored through naive-mean-field approximations~\cite{PRVBEM} and Approximate Message Passing (AMP)~\cite{PRGAMP}, the convergence of these algorithms is limited to random measurement matrix with i.i.d. sub-Gaussian entries. 
The great advantage of VAMP is its ability to achieve robust convergence across a broad class of matrices---right-orthogonally invariant random matrices---when the singular value decomposition (SVD) of the measurement matrix is accessible.
Notably, since the columns of $\mathsf{A}_{\text{ptycho}}$ are orthogonal, SVD adds no computational overhead for this matrix.

To enhance the stability of VAMP when applied to matrices that deviate from the random-matrix model, damping techniques~\cite{MRI_damping,deepEC} are commonly employed, albeit at the expense of a slower convergence.
Our prior work~\cite{StochasticVAMP} introduced a novel strategy to accelerate the convergence of VAMP, termed Stochastic-VAMP, derived from Belief Propagation on a tree graphical model.
In the problem setting of ptychographic reconstruction, Stochastic-VAMP can be viewed as a bayesian counterpart of PIE algorithms.

For blind ptychographic reconstruction, we combine the Expectation-Maximization (EM) algorithm~\cite{EM} with Stochastic-VAMP to jointly estimate the object $\bm{O}$ and the probe $\bm{P}$.
This method is an instance of Bad-VAMP~\cite{BadVAMP,BadGVAMP}, a framework for bilinear recovery problems.
The subsequent subsections describe the derivation of our algorithm, coined Ptycho-EP, using the notations established in Section II.

\subsection{Review on Belief Propagation}

We provide a concise review of the update rules in BP, using simple graphical models as examples.
Within this framework, two types of messages---"variable to factor" and "factor to variable"---are transmitted along the edges of a factor graph.
Figure 5 illustrates components of a factor graph, where $f_1, f_2,\cdots$ represent the factor nodes and $x_1,x_2,\cdots$ denote the variable nodes.
As depicted in Fig. 5. (a), when $x_1$ is connected to factor nodes $f_1, f_2, f_3$ ,  the "factor to variable" messages $m_{f_1 \rightarrow x_1}$ and $m_{f_2 \rightarrow x_1}$, along with "variable to factor" message $m_{x_1 \rightarrow f_3}$, are probabilistic distributions over $x_1$.
Given the messages incoming to variable $x_1$, the output from $x_1$ is given as follows:
\begin{align}
  m_{x_1 \rightarrow f_3}(x_1) \propto \prod_{i = 1}^2 m_{f_i \rightarrow x_1}(x_1)
\end{align}
The belief $b(x_1)$ is a probabilistic distribution given as
\begin{align}
  b(x_1) \propto \prod_{i=1}^3  m_{f_i \rightarrow x_1}(x_1)
\end{align}
which serves as an estimate of the marginal distribution of $x_1$.
From Eqs. (13) and (14), we have $m_{x_1 \rightarrow f_3}(x_1) \propto \frac{b(x_1)}{m_{f_3 \rightarrow x_1}(x_1)}$.
On the other hand, in Fig. 5. (b), messages arriving at the factor node $f_3$ is processed using integral:
\begin{equation}
  \begin{split}
  m_{f_3 \rightarrow x_1}(x_1) &\propto \\
  &\int f_3(x_1, x_2, x_3) \prod_{i = 2}^3 m_{x_i \rightarrow f_3}(x_i) dx_i
  \end{split}
\end{equation}

The sum-product algorithm updates "variable to factor" and "factor to variable" messages according to Eqs. (13) and Eq. (15), yielding the estimated marginal distribution from Eq. (14).
When the factor graph has no loops, this algorithm converges within a finite number of iterations, and the BP estimate of marginal is exact~\cite{mezard_information_2009}.
Given a subgraph $G$ in a factor graph as in Fig. 5. (c), the joint distribution of the variables included in $G$ is estimated as follows:
\begin{equation}
  \begin{split}
  b(x_1, x_2, x_3) &\propto f(x_1, x_2, x_3) \times\\
  &\prod_{m \in \mathcal{M}_G} m(x_1, x_2, x_3)
  \end{split}
\end{equation}
wherein $\mathcal{M}_G$ is the set of "factor to variable" messages incoming to the subgraph $G$. We refer to $b(x_1, x_2, x_3)$ as the joint belief on the variable nodes $x_1, x_2$, and $x_3$.

\begin{figure}[htb]
  \begin{minipage}[b]{1.0\linewidth}
    \centering
    \centerline{\includegraphics[width=8.5cm]{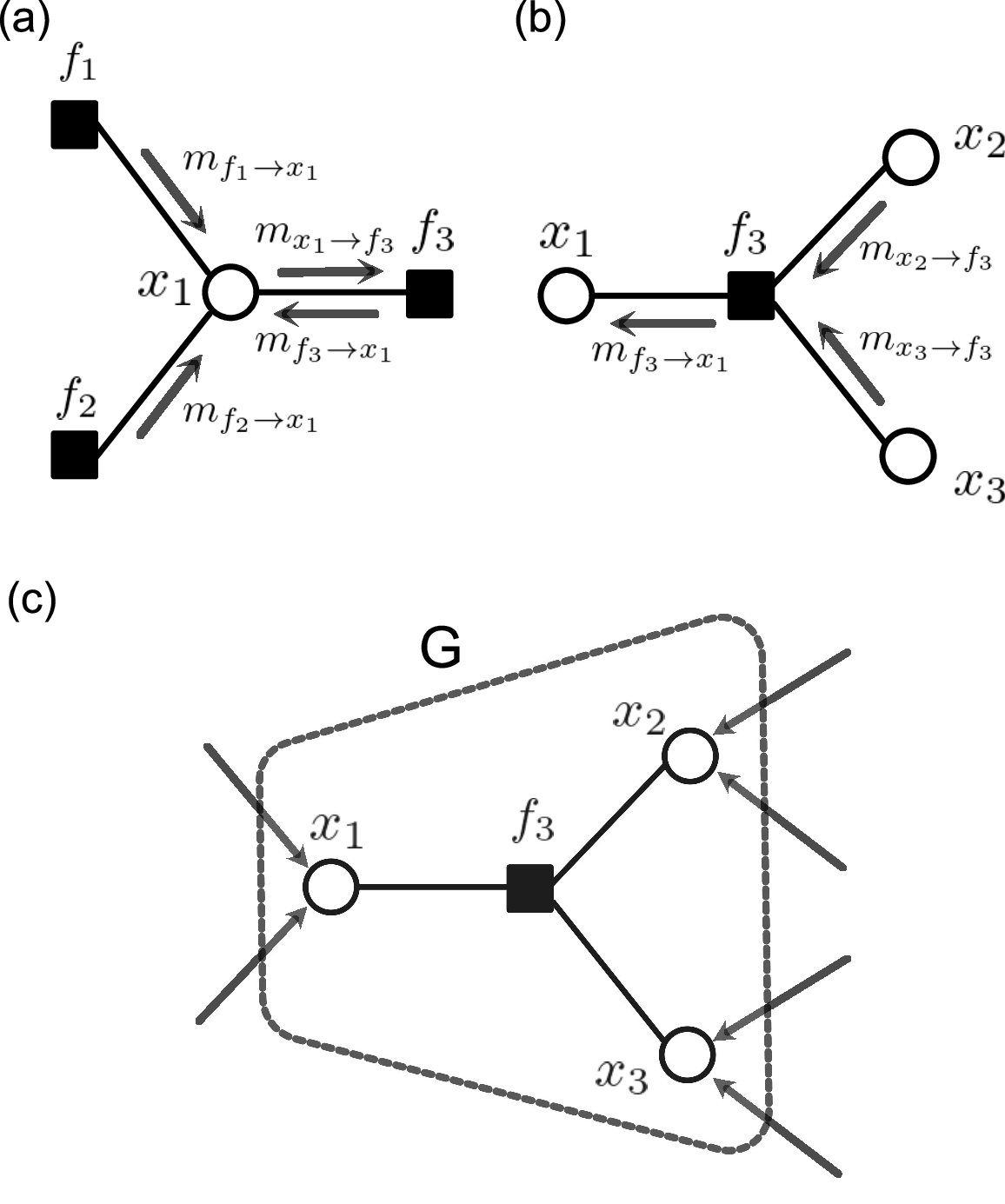}}
    \caption{Factor graphs to illustrate the update rules of Belief Propagation. (a) Massaging through a variable node. (b) Messaging through a factor node. (c) Joint distribution of variables within a subgraph $G$.}\medskip
  \end{minipage}
\end{figure}

In the derivation of Ptycho-EP, we assume messages are complex Gaussian distributions, which is a special case of Expectation Propagation (EP)~\cite{ExpectationPropagation}.
To describe the approximation that we will employ with EP, we define an operator $\text{Proj}[\cdots]$ which maps a probabilistic distribution on $\mathbb{C}^N$, denoted as  $p(\bm{x})$, to a Gaussian distribution given as
\begin{align}
  \text{Proj}[p] = \mathcal{CN}(\mathbb{E}(\bm{x}), \text{diag}[\text{Cov}(\bm{x})])
\end{align}
Here, $\mathbb{E}(\bm{x})$ and $\text{Cov}(\bm{x})$ denote the expectation and covariance matrix of $p(\bm{x})$.
This operator is employed to approximate messages by Gaussian, which is called the Expectation-Consistent (EC) approximation with vector-valued diagonalization~\cite{EC}.

\begin{figure*}[t]
  \begin{minipage}[b]{1.0\linewidth}
    \centering
    \centerline{\includegraphics[width=6.5cm, angle = 90]{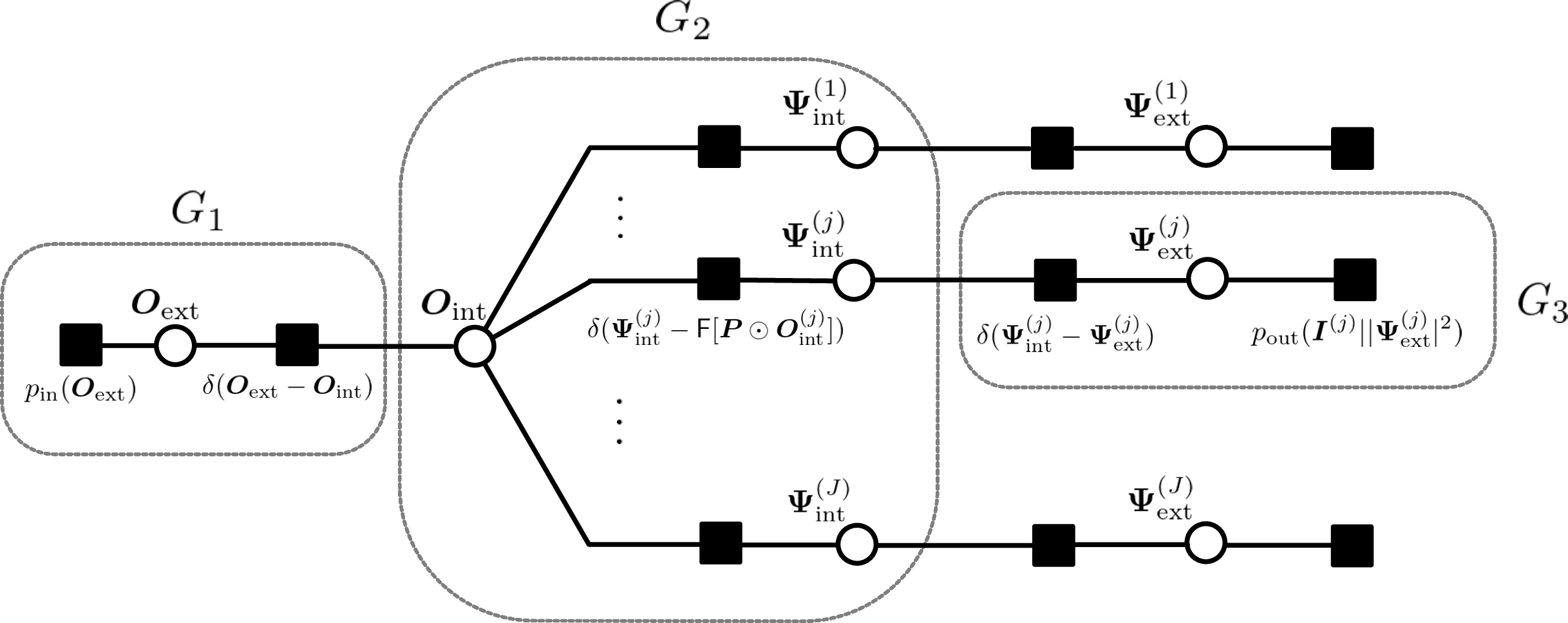}}
    \caption{An extended factor graph which is equivalent to the original graph in Fig. 4.}\medskip
  \end{minipage}
\end{figure*}

\subsection{Fixed point equations}
Following the derivation of VAMP~\cite{VAMP}, we introduce an equivalent model of Eq. (9) by replicating variables $\bm{O}, \bm{\Psi}^{(j)} \ (j = 1,..., J)$ into $\bm{O}_{\text{int}}, \bm{O}_{\text{ext}}, \bm{\Psi}^{(j)}_{\text{int}}, \bm{\Psi}^{(j)}_{\text{ext}} $, and connect the pairs by Dirac delta function nodes $\delta(\bm{O}_{\text{int}} - \bm{O}_{\text{ext}})$ and $\delta(\bm{\Psi}^{(j)}_{\text{int}} - \bm{\Psi}^{(j)}_{\text{ext}})$.
The corresponding factor graph is illustrated in Fig. 6.
We derive the fixed point equations through the EP in the subgraphs $G_1, G_2$, and $G_3$.

(1) \underline{Subgraph $G_1$}\ First, we assume that the message sent from $\delta(\bm{O}_{\text{ext}} - \bm{O}_{\text{int}})$ node to $\bm{O}_{\text{ext}}$ node is a Gaussian distribution $\mathcal{CN}(\bm{O}_{\text{ext}} ; \widetilde{\bm{O}}_{\text{ext}}, \mathsf{Diag}(\widetilde{\bm{\Gamma}}^{-1}_{\text{O,ext}}))$.
Here, $\widetilde{\bm{O}}_{\text{ext}} \in \mathbb{C}^N$ is the mean and $\widetilde{\bm{\Gamma}}_{\text{O,ext}} \in \mathbb{R}_+^N$ denotes the precision.
This creates a belief on $\bm{O}_{\text{ext}}$ node, given as $b(\bm{O}_{\text{ext}}) \propto p_{\text{in}}(\bm{O}_{\text{ext}})\times \mathcal{CN}(\bm{O}_{\text{ext}} ; \widetilde{\bm{O}}_{\text{ext}}, \mathsf{Diag}(\widetilde{\bm{\Gamma}}^{-1}_{\text{O,ext}}))$.
We introduce the denoising functions $\bm{g}_{\text{in}}(\widetilde{\bm{O}}_{\text{ext}}, \widetilde{\bm{\Gamma}}_{\text{O,ext}})$ and $\bm{g}'_{\text{in}}(\widetilde{\bm{O}}_{\text{ext}}, \widetilde{\bm{\Gamma}}_{\text{O,ext}})$, defined by the following projection:
\begin{align}
  \text{Proj}[b(\bm{O}_{\text{ext}})] &= \mathcal{CN}(\bm{O}_{\text{ext}} ; \widehat{\bm{O}}_{\text{ext}}, \mathsf{Diag}(\hat{\bm{\Gamma}}^{-1}_{\text{O,ext}}))\nonumber \\
  \widehat{\bm{O}}_{\text{ext}} &= \bm{g}_{\text{in}}(\widetilde{\bm{O}}_{\text{ext}}, \widetilde{\bm{\Gamma}}_{\text{O,ext}})\\
  \widehat{\bm{\Gamma}}^{-1}_{\text{O,ext}} &= \bm{g}'_{\text{in}}(\widetilde{\bm{O}}_{\text{ext}}, \widetilde{\bm{\Gamma}}_{\text{O,ext}})
\end{align}
For instance, in the case of Gaussian prior $p_{\text{in}} = \mathcal{CN}(\bm{0}, \mathsf{I}_N)$, $\bm{g}_{\text{in}}(\widetilde{\bm{O}}_{\text{ext}}, \widetilde{\bm{\Gamma}}_{\text{O,ext}}) = (\widetilde{\bm{\Gamma}}_{\text{O,ext}} \odot \widetilde{\bm{O}}_{\text{ext}})/(\bm{1}+\widetilde{\bm{\Gamma}}_{\text{O,ext}})$ and $\bm{g}'_{\text{in}}(\widetilde{\bm{O}}_{\text{ext}}, \widetilde{\bm{\Gamma}}_{\text{O,ext}}) = (\bm{1}+\widetilde{\bm{\Gamma}}_{\text{O,ext}})^{-1}$.
The denoising functions for BG prior in Eq. (7) is derived in ~\cite{PRGAMP}. For readability, the explicit form is written using our notations in Appendix A.
The variable node $\bm{O}_{\text{ext}}$, then, sends the message $\mathcal{CN}(\bm{O}_{\text{ext}} ; \widetilde{\bm{O}}_{\text{int}}, \mathsf{Diag}(\widetilde{\bm{\Gamma}}^{-1}_{\text{O,int}}))  \propto \text{Proj}[b(\bm{O}_{\text{ext}})]/\mathcal{CN}(\bm{O}_{\text{ext}} ; \widetilde{\bm{O}}_{\text{ext}}, \mathsf{Diag}(\widetilde{\bm{\Gamma}}^{-1}_{\text{O,ext}})) $ back to the delta function node $\delta(\bm{O}_{\text{ext}} - \bm{O}_{\text{ext}})$, where
\begin{align}
  \widetilde{\bm{O}}_{\text{int}} &= \frac{\widehat{\bm{\Gamma}}_{\text{O,ext}} \odot \widehat{\bm{O}}_{\text{ext}} - \widetilde{\bm{\Gamma}}_{\text{O,ext}} \odot \widetilde{\bm{O}}_{\text{ext}}}{\widehat{\bm{\Gamma}}_{\text{O,ext}} - \widetilde{\bm{\Gamma}}_{\text{O,ext}}}\\
  \widetilde{\bm{\Gamma}}_{\text{O,int}} &= \widehat{\bm{\Gamma}}_{\text{O,ext}} - \widetilde{\bm{\Gamma}}_{\text{O,ext}}
\end{align}
The message $\mathcal{CN}(\bm{O}_{\text{int}} ; \widetilde{\bm{O}}_{\text{int}}, \mathsf{Diag}(\widetilde{\bm{\Gamma}}^{-1}_{\text{O,int}}))$ is propagated to the subgraph $G_2$, after traversing the delta function node.

(2) \underline{Subgraph $G_2$}
We denote by $\mathcal{CN}(\bm{O}_{\text{int}} ; \widetilde{\bm{O}}_{\text{int}}, \mathsf{Diag}(\widetilde{\bm{\Gamma}}^{-1}_{\text{O,int}}))$ the message sent from $\delta(\bm{O}_{\text{ext}} - \bm{O}_{\text{int}})$ node to $G_2$, and by $\mathcal{CN}(\bm{\Psi}_{\text{int}}^{(j)} ; \widetilde{\bm{\Psi}}^{(j)}_{\text{int}}, \widetilde{{\Gamma}}^{(j) -1}_{\Psi,\text{int}} \mathsf{I}_M)$ the message sent from $\delta(\bm{\Psi}^{(j)}_{\text{ext}} - \bm{\Psi}^{(j)}_{\text{int}})$ node to $G_2$ ($j = 1,\cdots,J$).
Note that, while the individual components of $\widetilde{\bm{\Gamma}}_{\text{int}}$ is allowed to have a distinct value, we assume that the covariance matrix of $\bm{\Psi}^{(j)}_{\text{int}}$ to be a scalar.
The joint belief on variable nodes $\bm{O}_{\text{int}}, \bm{\Psi}^{(j)}_{\text{int}} (j = 1,\cdots J)$ is given as
\begin{equation}
\begin{split}
  b(\bm{O}_{\text{int}}, &\bm{\Psi}^{(1)}_{\text{int}},\cdots,\bm{\Psi}^{(J)}_{\text{int}}) \\
  &\propto \mathcal{CN}(\bm{O}_{\text{int}} ; \widetilde{\bm{O}}_{\text{int}}, \mathsf{Diag}(\widetilde{\bm{\Gamma}}^{-1}_{\text{O,int}})) \times \\
  & \prod_{j=1}^J \delta (\bm{\Psi}^{(j)}_{\text{int}} - \mathsf{F}[\bm{P}\odot\bm{O}^{(j)}_{\text{int}}]) \times \\
  & \prod_{j=1}^J \mathcal{CN}(\bm{\Psi}_{\text{int}}^{(j)} ; \widetilde{\bm{\Psi}}^{(j)}_{\text{int}}, \widetilde{{\Gamma}}^{(j) -1}_{\Psi,\text{int}} \mathsf{I}_M)\nonumber
\end{split}
\end{equation}
where $\bm{O}^{(j)}_{\text{int}} \triangleq \mathsf{S}^{(j)} \bm{O}_{\text{int}}$.
The marginal of $\bm{O}_{\text{int}}$ is expressed as $b(\bm{O}_{\text{int}}) = \mathcal{CN}(\bm{O}_{\text{int}} ; \widehat{\bm{O}}_{\text{int}}, \mathsf{Diag}(\widehat{\bm{\Gamma}}^{-1}_{\text{O,int}}))$ with
\begin{align}
  \widehat{\bm{\Gamma}}_{\text{O,int}} &= \widetilde{\bm{\Gamma}}_{\text{O,int}} + \sum_{j=1}^J \widetilde{{\Gamma}}^{(j)}_{\Psi,\text{int}} \mathsf{S}^{(j)T}|\bm{P}|^2\\
  \widehat{\bm{O}}_{\text{int}} &= \frac{\widehat{\bm{\Lambda}}_{\text{O,int}}}{\widehat{\bm{\Gamma}}_{\text{O,int}}}
\end{align}
Here, the auxiliary variable $\widehat{\bm{\Lambda}}_{\text{O,int}}$ is defined by
\begin{equation}
\begin{split}
    \widehat{\bm{\Lambda}}_{\text{O,int}} = &\widetilde{\bm{\Gamma}}_{\text{O,int}} \odot \widetilde{\bm{O}}_{\text{int}} \\
    &+ \sum_{j=1}^J \widetilde{{\Gamma}}^{(j)}_{\Psi,\text{int}} \mathsf{S}^{(j)T}[\bm{P}^* \odot \mathsf{F}^H \widetilde{\bm{\Psi}}^{(j)}_{\text{int}}]
  \end{split}
\end{equation}
Having computed $\widehat{\bm{O}}_{\text{int}}$ and $\widehat{\bm{\Gamma}}_{\text{O,int}}$, the belief on $\bm{\Psi}^{(j)}_{\text{int}}$ node is simply given as
\begin{equation}
  \begin{split}
  b(&\bm{\Psi}^{(j)}_{\text{int}}) =\\\nonumber
   &\mathcal{CN}(\bm{\Psi}^{(j)}_{\text{int}} ; \mathsf{F}[\bm{P}\odot\widehat{\bm{O}}^{(j)}_{\text{int}}], \mathsf{F}\mathsf{Diag}(|\bm{P}|^2 \odot \widehat{\bm{\Gamma}}^{(j)-1}_{\text{O,int}})\mathsf{F}^H)
\end{split}
\end{equation}
wherein $\widehat{\bm{O}}^{(j)}_{\text{int}} \triangleq \mathsf{S}^{(j)}\widehat{\bm{O}}_{\text{int}}$ and $\widehat{\bm{\Gamma}}^{(j)}_{\text{O,int}} \triangleq \mathsf{S}^{(j)}\widehat{\bm{\Gamma}}_{\text{O,int}}$.
Neglecting the off-diagonal elements of covariance, the marginal is approximated by $\text{Proj}[b(\bm{\Psi}^{(j)}_{\text{int}})] = \mathcal{CN}(\bm{\Psi}^{(j)}_{\text{int}} ; \widehat{\bm{\Psi}}^{(j)}_{\text{int}}, \widehat{\Gamma}^{(j)-1}_{\Psi,\text{int}} I_M)$ with
\begin{align}
  \widehat{\bm{\Psi}}^{(j)}_{\text{int}} &= \mathsf{F}[\bm{P}\odot\widehat{\bm{O}}^{(j)}_{\text{int}}]\\
  \widehat{\Gamma}^{(j)}_{\Psi,\text{int}} &= \left(\frac{1}{M} \text{tr}(|\bm{P}|^2 \odot \widehat{\bm{\Gamma}}^{(j)-1}_{\text{O,int}})\right)^{-1}
\end{align}
By the update rules of BP, subgraph $G_2$ transmits to $G_1$ the massage $\mathcal{CN}(\bm{O}_{\text{ext}} ; \widetilde{\bm{O}}_{\text{ext}}, \mathsf{Diag}(\widetilde{\bm{\Gamma}}^{-1}_{\text{O,ext}}))$ with
\begin{align}
  \widetilde{\bm{O}}_{\text{ext}} &= \frac{\widehat{\bm{\Gamma}}_{\text{O,int}} \odot \widehat{\bm{O}}_{\text{int}} - \widetilde{\bm{\Gamma}}_{\text{O,int}} \odot \widetilde{\bm{O}}_{\text{int}}}{\widehat{\bm{\Gamma}}_{\text{O,int}} - \widetilde{\bm{\Gamma}}_{\text{O,int}}}\\
  \widetilde{\bm{\Gamma}}_{\text{O,ext}} &= \widehat{\bm{\Gamma}}_{\text{O,int}} - \widetilde{\bm{\Gamma}}_{\text{O,int}}
\end{align}
Similarly, $G_2$ feeds $G_3$ with the message $\mathcal{CN}(\bm{\Psi}^{(j)}_{\text{ext}}; \widetilde{\bm{\Psi}}^{(j)}_{\text{ext}}, \widetilde{\Gamma}^{(j)-1}_{\Psi, \text{ext}} \mathsf{I}_M)$, where
\begin{align}
  \widetilde{\bm{\Psi}}^{(j)}_{\text{ext}} &= \frac{\widehat{\Gamma}^{(j)}_{\Psi,\text{int}} \widehat{\bm{\Psi}}^{(j)}_{\text{int}} - \widetilde{{\Gamma}}^{(j)}_{\Psi,\text{int}} \widetilde{\bm{\Psi}}^{(j)}_{\text{int}}}{\widehat{\Gamma}^{(j)}_{\Psi,\text{int}} - \widetilde{{\Gamma}}^{(j)}_{\Psi,\text{int}}}\\
  \widetilde{\Gamma}^{(j)}_{\Psi, \text{ext}} &= \widehat{\Gamma}^{(j)}_{\Psi,\text{int}} - \widetilde{{\Gamma}}^{(j)}_{\Psi,\text{int}}
\end{align}

(3) \underline{Subgraph $G_3$} EP updates in $G_3$ is analogous to those in $G_1$, except that the variance of $\bm{\Psi}^{(j)}$ ($j = 1,\cdots,J$) is constrained to be a scalar.
The local belief on $\bm{\Psi}^{(j)}_{\text{ext}}$ node is approximated by Gaussian $\mathcal{CN}(\bm{\Psi}^{(j)}_{\text{ext}}; \widehat{\bm{\Psi}}^{(j)}_{\text{ext}}, \widehat{\Gamma}^{(j)-1}_{\Psi,\text{ext}} I_M)$, where
\begin{align}
  \widehat{\bm{\Psi}}^{(j)}_{\text{ext}} &= \bm{g}_{\text{out}}(\widetilde{\bm{\Psi}}^{(j)}_{\text{ext}}, \widetilde{\Gamma}^{(j)}_{\Psi, \text{ext}}; \bm{I}^{(j)})\\
  \widehat{\Gamma}^{(j)-1}_{\Psi,\text{ext}} &= \frac{1}{M} \left( \text{tr}(\bm{g}'_{\text{out}}(\widetilde{\bm{\Psi}}^{(j)}_{\text{ext}}, \widetilde{\Gamma}^{(j)}_{\Psi, \text{ext}}; \bm{I}^{(j)}))\right)
\end{align}
The data-dependent denoiser $\bm{g}_{\text{out}}(\widetilde{\bm{\Psi}}^{(j)}_{\text{ext}}, \widetilde{\Gamma}^{(j)}_{\Psi, \text{ext}}; \bm{I}^{(j)})$ and $\bm{g}'_{\text{out}}(\widetilde{\bm{\Psi}}^{(j)}_{\text{ext}}, \widetilde{\Gamma}^{(j)}_{\Psi, \text{ext}}; \bm{I}^{(j)})$ are determined by the noise model.
An explicit form of denoising function corresponding to the model defined in Eq. (6), which we will employ throughout this paper, is derived in ~\cite{deepEC} and provided in Appendix. A for readability.
This unit propagates the message $\mathcal{CN}(\bm{\Psi}^{(j)}_{\text{int}}; \widetilde{\bm{\Psi}}^{(j)}_{\text{int}}, \widetilde{\Gamma}^{(j)-1}_{\Psi, \text{int}} \mathsf{I}_M)$ to subgraph $G_2$, where
\begin{align}
  \widetilde{\bm{\Psi}}^{(j)}_{\text{int}} &= \frac{\widehat{\Gamma}^{(j)}_{\Psi,\text{ext}} \widehat{\bm{\Psi}}^{(j)}_{\text{ext}} - \widetilde{\Gamma}^{(j)}_{\Psi, \text{ext}} \widetilde{\bm{\Psi}}^{(j)}_{\text{ext}}}{\widehat{\Gamma}^{(j)}_{\Psi,\text{ext}} - \widetilde{\Gamma}^{(j)}_{\Psi, \text{ext}}}\\
  \widetilde{\Gamma}^{(j)}_{\Psi, \text{int}} &= \widehat{\Gamma}^{(j)}_{\Psi,\text{ext}} - \widetilde{\Gamma}^{(j)}_{\Psi, \text{ext}}
\end{align}
Our algorithm searches the fixed points of the equation system Eqs. (18)-(34), and outputs the fixed point value of $\widehat{\bm{O}}_{\text{int}}$ as an estimate of $\mathbb{E}(\bm{O} | \bm{I}^{(1)},\cdots,\bm{I}^{(J)})$.
Furthermore, $\widehat{\bm{\Gamma}}^{-1}_{O, \text{int}}$ serves as an uncertainty quantification of the reconstruction by Ptycho-EP.
Note that, at the fixed point, $\widehat{\bm{\Gamma}}_{O, \text{int}} = \widehat{\bm{\Gamma}}_{O, \text{ext}}$ from Eqs. (21) and (28), and when $\widehat{\bm{\Gamma}}_{O, \text{int}} > 0$, we have $\widehat{\bm{O}}_{\text{int}} = \widehat{\bm{O}}_{\text{ext}}$ from Eqs. (20) and (27).

\subsection{Stochastic VAMP}
We introduce an efficient iterative scheme, coined Stochastic VAMP, that finds fixed points of the equation system derived in the previous section.
First, we highlight the three components in our algorithm : \\

\textbf{Belief Update}: \\
Input : $(\widetilde{\bm{O}}_{\text{int}}, \widetilde{\bm{\Gamma}}_{O,\text{int}})$ and all of the $J$ messages $(\widetilde{\bm{\Psi}}^{(1)}_{\text{int}}, \widetilde{\Gamma}^{(1)}_{\Psi,\text{int}}), \cdots, (\widetilde{\bm{\Psi}}^{(J)}_{\text{int}}, \widetilde{\Gamma}^{(J)}_{\Psi,\text{int}})$ \\
Output : $\widehat{\bm{O}}_{\text{int}}$ and $\widehat{\bm{\Gamma}}_{O,\text{int}}$ computed by Eqs.(22)-(24)\\

\textbf{Message Update by Data}:\\
Input : $(\widehat{\bm{O}}_{\text{int}}, \widehat{\bm{\Gamma}}_{O,\text{int}})$, $j \in \{1,\cdots,J\}$ and $(\widetilde{\bm{\Psi}}^{(j)}_{\text{int}}, \widetilde{\Gamma}^{(j)}_{\Psi,\text{int}})$\\
Output : New value of $(\widetilde{\bm{\Psi}}^{(j)}_{\text{int}}, \widetilde{\Gamma}^{(j)}_{\Psi,\text{int}})$ computed by sequentially applying Eqs. (25)-(26),(29)-(34)\\

\textbf{Message Update by Prior}:\\
Input : $(\widehat{\bm{O}}_{\text{int}}, \widehat{\bm{\Gamma}}_{O,\text{int}})$ and $(\widetilde{\bm{O}}_{\text{int}}, \widetilde{\bm{\Gamma}}_{O,\text{int}})$\\
Output : New value of $(\widetilde{\bm{O}}_{\text{int}}, \widetilde{\bm{\Gamma}}_{O,\text{int}})$ computed by sequentially applying Eqs. (27)-(28),(18)-(21)\\

For clarity, we focus on the case $p_{\text{in}}(\bm{O}) = \mathcal{CN}(\bm{O}; \bm{0}, \mathsf{I}_N)$, where \textit{Message Update by Prior} module does nothing but yielding $\widetilde{\bm{O}}_{\text{int}} = \bm{0}, \widetilde{\bm{\Gamma}}_{O,\text{int}} = \bm{1}$ regardless of the input.
Under this setting, there are two iterative schemes to combine \textit{Message Update by Data} and \textit{Belief Update}, outlined in Algorithm 1 and Algorithm 2.

\begin{algorithm}[htb]
  \caption{Parallel Scheme}\label{alg:cap}
  \begin{algorithmic}[1]
    \Procedure{Update}{$\widetilde{\bm{O}}_{\text{int}}, \widetilde{\bm{\Gamma}}_{O,\text{int}}$, $\{\widetilde{\bm{\Psi}}^{(j)}_{\text{int}}, \widetilde{\Gamma}^{(j)}_{\Psi,\text{int}}\}^J_{j=1}$}
    \State Compute $\widehat{\bm{O}}_{\text{int}}$ and $\widehat{\bm{\Gamma}}_{O,\text{int}}$ by \textit{Belief Update}
    \For{$j = 1,\cdots,J$}
    \State Update $\widetilde{\bm{\Psi}}^{(j)}_{\text{int}}$ and $\widetilde{\Gamma}^{(j)}_{\Psi,\text{int}}$ by \textit{Message Update} 
    \Statex \quad \quad \quad \textit{by Data} with $j$-th data item $\bm{I}^{(j)}$
    \EndFor
    \EndProcedure
  \end{algorithmic}  
\end{algorithm}

\begin{algorithm}[htb]
  \caption{Sequential Scheme}\label{alg:cap}
  \begin{algorithmic}[1]
    \Procedure{Update}{$\widetilde{\bm{O}}_{\text{int}}, \widetilde{\bm{\Gamma}}_{O,\text{int}}$, $\{\widetilde{\bm{\Psi}}^{(j)}_{\text{int}}, \widetilde{\Gamma}^{(j)}_{\Psi,\text{int}}\}^J_{j=1}$}
    \For{$j = 1,\cdots,J$}
    \State Compute $\widehat{\bm{O}}_{\text{int}}$ and $\widehat{\bm{\Gamma}}_{O,\text{int}}$ by \textit{Belief Update}
    \State Update $\widetilde{\bm{\Psi}}^{(j)}_{\text{int}}$ and $\widetilde{\Gamma}^{(j)}_{\Psi,\text{int}}$ by \textit{Message Update} 
    \Statex \quad \quad \quad \textit{by Data} with $j$-th data item $\bm{I}^{(j)}$
     \EndFor
     \EndProcedure
  \end{algorithmic}  
\end{algorithm}

Since \textit{Message Update by Data} requires looking at only one of the $J$ data items, the $j$-loop in Algorithm 1 can be executed in parallel.
Conversely, in the sequential scheme, \textit{Belief Update} is called each time one of the $J$ messages is updated to accelerate convergence.
Note that, while the \textit{Belief Update} in Algorithm 1 involves the summation of $J$ variables, this can be avoided in the sequential scheme by using the following "incremental" update rule, instead of Eqs.(22) and (24).
\begin{align}
  \widehat{\bm{\Gamma}}^{\text{new}}_{\text{O,int}} &= \widehat{\bm{\Gamma}}^{\text{old}}_{\text{O,int}} + (\widetilde{{\Gamma}}^{(j), \text{new}}_{\Psi,\text{int}} - \widetilde{{\Gamma}}^{(j), \text{old}}_{\Psi,\text{int}}) \mathsf{S}^{(j)T}|\bm{P}|^2
\end{align}
\begin{equation}
    \begin{split}
      \widehat{\bm{\Lambda}}^{\text{new}}_{\text{O,int}} = \widehat{\bm{\Lambda}}^{\text{old}}_{\text{O,int}} &+ \widetilde{{\Gamma}}^{(j),\text{new}}_{\Psi,\text{int}} \mathsf{S}^{(j)T}[\bm{P}^* \odot \mathsf{F}^H \widetilde{\bm{\Psi}}^{(j),\text{new}}_{\text{int}}]\\
    &- \widetilde{{\Gamma}}^{(j),\text{old}}_{\Psi,\text{int}} \mathsf{S}^{(j)T}[\bm{P}^* \odot \mathsf{F}^H \widetilde{\bm{\Psi}}^{(j),\text{old}}_{\text{int}}]
  \end{split}
\end{equation}
wherein $\widehat{\bm{\Lambda}}^{\text{old}}_{\text{O,int}}$, $\widehat{\bm{\Gamma}}^{\text{old}}_{\text{O,int}}$, and $\widetilde{\bm{\Psi}}^{(j),\text{old}}_{\text{int}}$ are the previous values of $\widehat{\bm{\Lambda}}_{\text{O,int}}$, $\widehat{\bm{\Gamma}}_{\text{O,int}}$, and $\widetilde{\bm{\Psi}}^{(j)}_{\text{int}}$, which should be retained on the memory.
$\widetilde{{\Gamma}}^{(j),\text{new}}_{\Psi,\text{int}}$ is the new value of $\widetilde{{\Gamma}}^{(j)}_{\Psi,\text{int}}$ computed by \textit{Message Update by Data} component, and the left hand sides of Eqs. (35)-(36) are the outputs from \textit{Belief Update}.

The overall computational costs of Algorithm 1 and 2 are primarily dominated by $J$ matrix-multiplications by $\mathsf{F}$ and $\mathsf{F}^H$, which is a common complexity of ptychographic reconstruction algorithms such as ePIE.
We refer to the sequential approach as \textit{Stochastic VAMP}, whose rapid convergence will be validated in section IV.

\subsection{Algorithmic details}
\textit{Damping}---Instead of using the outputs from \textit{Message Update by Data} module directly, we update $(\widetilde{\bm{\Psi}}^{(j)}_{\text{int}}, \widetilde{\Gamma}^{(j)}_{\Psi,\text{int}})$ by the following:
\begin{equation}
\begin{split}
  &\widetilde{\bm{\Psi}}^{(j)}_{\text{int}} \leftarrow \mu \widetilde{\bm{\Psi}}^{(j), \text{raw}}_{\text{int}} + (1 - \mu) \widetilde{\bm{\Psi}}^{(j), \text{old}}_{\text{int}}\\ 
  &\widetilde{\Gamma}^{(j)}_{\Psi,\text{int}} \leftarrow \left\{ \mu (\widetilde{\Gamma}^{(j),\text{raw}}_{\Psi,\text{int}})^{-\frac{1}{2}} + (1 - \mu)  (\widetilde{\Gamma}^{(j),\text{old}}_{\Psi,\text{int}})^{-\frac{1}{2}}\right\}^2 
\end{split}
\end{equation}
Here, $\mu \in (0,1)$ denotes the damping constant, and $(\widetilde{\bm{\Psi}}^{(j), \text{raw}}_{\text{int}}, \widetilde{\Gamma}^{(j),\text{raw}}_{\Psi,\text{int}})$ is the direct output from \textit{Message Update by Data} component.
This damping strategy, proposed in ~\cite{MRI_damping}, is designed to promote the convergence of VAMP, albeit at the expense of slower updates.
Fortunately, our numerical experience suggests that the convergence of Stochastic VAMP remains stable with damping constant close to 1, when compared with the parallel scheme.

\textit{Probe Retrieval by EM}---Thus far, we have assumed that the probe $\bm{P}$ is known a priori.
However, it can be estimated from the diffraction dataset, by incorporating the EM algorithm with Ptycho-EP, as detailed in Appendix B.
In the adaptive EM framework~\cite{BadVAMP,BadGVAMP}, $\bm{P}$ and the precision parameters $\widetilde{\Gamma}^{(j)}_{\Psi,\text{int}}\ (j = 1,\cdots,J)$ are alternately updated several times.
The overall algorithm is summarized in Algorithm 3.

\begin{algorithm}[htb]
  \caption{Ptycho-EP with EM}\label{alg:cap}
  \begin{algorithmic}[1]
    \Require Dataset $\{\mathsf{S}^{(j)}, \bm{I}^{(j)}\}_{j=1}^J$, Damping constant $\mu$, Maximum iteration number $T$, per-iteration number of EM updates $\tau$
    \State Initialize $\widetilde{\bm{O}}_{\text{int}}, \widetilde{\bm{\Gamma}}_{O,\text{int}}$, $\{\widetilde{\bm{\Psi}}^{(j)}_{\text{int}}, \widetilde{\Gamma}^{(j)}_{\Psi,\text{int}}\}^J_{j=1}$, and $\bm{P}$ 
    \For{$t = 1,\cdots,T$}
    \Statex \quad \quad module
     \For{$j = 1,\cdots,J$}
        \State Compute $\widehat{\bm{O}}_{\text{int}}$ and $\widehat{\bm{\Gamma}}_{O,\text{int}}$ by \textit{Belief Update} module
        \State Compute $(\widetilde{\bm{\Psi}}^{(j), \text{raw}}_{\text{int}}$ and $\widetilde{\Gamma}^{(j),\text{raw}}_{\Psi,\text{int}})$ by \textit{Message} 
        \Statex \quad \quad \quad \textit{Update by Data} module
        \State Update $(\widetilde{\bm{\Psi}}^{(j)}_{\text{int}}, \widetilde{\Gamma}^{(j)}_{\Psi,\text{int}})$ using Eq. (37)
    \EndFor
    \State Compute $\widehat{\bm{O}}_{\text{int}}$ and $\widehat{\bm{\Gamma}}_{O,\text{int}}$ by \textit{Belief Update} module
    \State Update $\widetilde{\bm{O}}_{\text{int}}$ and $\widetilde{\bm{\Gamma}}_{O,\text{int}}$ by \textit{Message Update by Prior}     
    \For{$k = 1,\cdots,\tau$}
      \State Update $\bm{P}$ and $\widetilde{\Gamma}^{(j)}_{\Psi,\text{int}}\ (j = 1,\cdots,J)$ by \textit{Adaptive} 
      \Statex \quad \quad \quad \textit{EM} module
    \EndFor
    \Statex \Comment Line 10-12 is omitted when $\bm{P}$ is known a priori
    \EndFor
  \Ensure  $\widehat{\bm{O}}_{\text{int}}$, $\widehat{\bm{\Gamma}}_{O,\text{int}}$, and $\bm{P}$
  \end{algorithmic}  
\end{algorithm}

\section{Computational results}
\subsection{Dataset and Algorithm Setup}

\begin{figure*}[t]
  \begin{minipage}[b]{1.0\linewidth}
    \centering
    \centerline{\includegraphics[width=5.5cm, angle = 90]{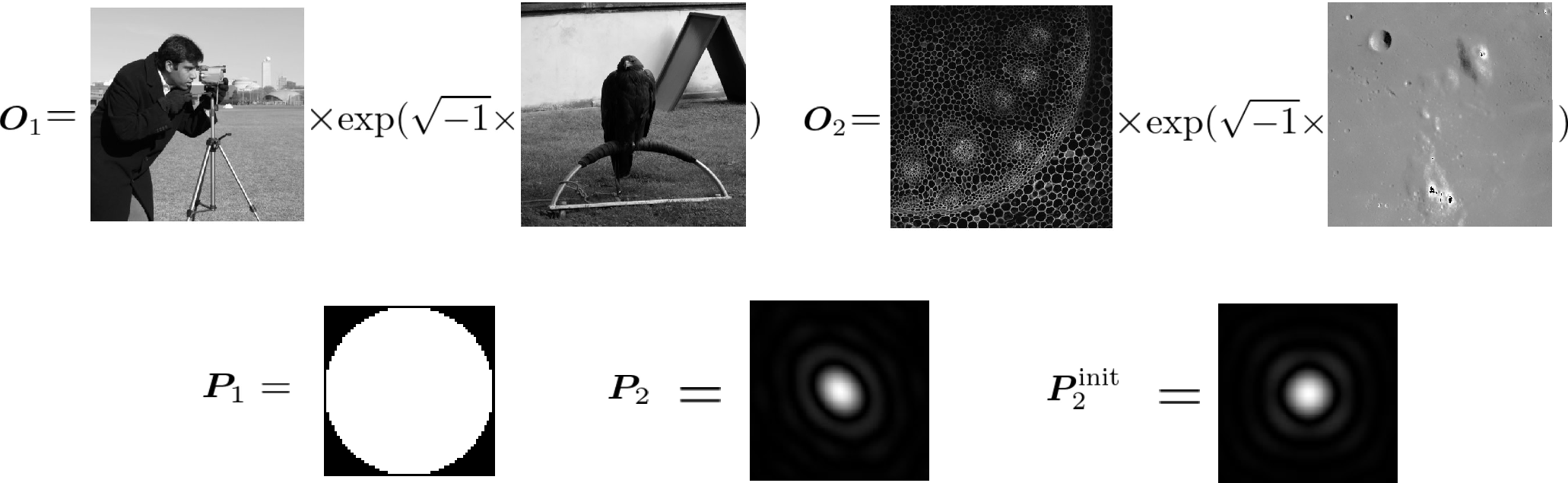}}
    \caption{(Upper) test data for complex object $\bm{O}_1 \in \mathbb{C}^{512 \times 512}$ and $\bm{O}_2 \mathbb{C}^{512 \times 512}$. The 60 \% of the pixels in $\bm{O}_2$ is zero. (Lower) test data for probe $\bm{P}_1 \in \mathbb{C}^{64 \times 64}$ and $\bm{P}_2,\bm{P}_2^{\text{init}}  \in \mathbb{C}^{128 \times 128}$. $\bm{P}_2^{\text{init}}$ is used as an initial estimate of $\bm{P}_2$ in blind ptychographic phase retrieval.}\medskip
  \end{minipage}
\end{figure*}

$512 \times 512$-pixel complex images $\bm{O}_1$ and $\bm{O}_2$ are synthesized with two pairs of real images from scikit-image dataset ~\cite{scikit-image}, as depicted in Fig.7.
The pixel-wise absolute value of $\bm{O}_1$ and $\bm{O}_2$ is distributed in the range $[0,1]$, while their complex angle lie in the interval $[0, \pi/2]$.
Furthermore, $\bm{O}_2$ is preprocessed so that $60\%$ of pixel values are zero, in order to evaluate the performance of Ptycho-EP with sparse prior.

In numerical experiments where the probe is assumed to be accurately known, we use the simple round aperture $\bm{P}_1 \in \mathbb{C}^{64 \times 64}$ as the probe.
For evaluating ptychographic reconstruction algorithms with probe retrieval, datasets are generated by using $\bm{P}_2 \in \mathbb{C}^{128 \times 128}$ as the true probe.
In these cases, $\bm{P}^{\text{init}}_2 \in \mathbb{C}^{128 \times 128}$ is provided to the algorithms as initial estimate of $\bm{P}_2$.

Throughout our simulations, 2-D DFT is employed as the propagation operator $\mathsf{F}$.
For sequential algorithms such as PIE and Ptycho-EP, the data items $\bm{I}^{(1)}, \cdots, \bm{I}^{(J)}$ are sorted by the distance between scan position and the center of the $512 \times 512$ pixel image.

For Ptycho-EP with Gaussian prior, the damping constant is set to $\mu = 0.9$, whereas $\mu = 0.95$ is used for Ptycho-EP with sparse prior. 
The number of EM updates is fixed at $\tau = 2$ for probe reconstruction.
Following the notation established in~\cite{rPIE}, the tuning parameters for algorithms within the PIE family are specified as follows: $\alpha = 0.1$ for PIE~\cite{APL_PIE}, $\alpha = \beta = 1$ for ePIE~\cite{ePIE}, and $\alpha = 0.1, \beta = 1$ for rPIE~\cite{rPIE}.
The probe update rule for rPIE is identical to that of ePIE, as suggested in~\cite{rPIE}.
For Difference Map algorithm~\cite{DifferenceMap}, a standard choice of tuning parameter $\beta = 1$ is employed ~\cite{ComputationalFramework}.
The initial guess of object is randomly drawn from $\mathcal{CN}(\bm{0}, I_N)$ for all of the algorithms.

\subsection{Noise Model and Error Metric}
The noise corrupted observation of $\sqrt{\bm{I}^{(j)}}\ (j = 1,\cdots,J)$ is generated by adding white Gaussian noise to the noiseless value of $\sqrt{\bm{I}^{(j)}}$, serving as an approximation of Poisson noise affecting $\bm{I}^{(j)}$.
This noise model corresponds to the likelihood in Eq.(6), and the noise level is controlled by varying the value of $\sigma$.
We define the signal-to-noise ratio (SNR) of the measurement by
\begin{align}
  \text{SNR}^{-1} = \frac{\sigma^2}{\frac{1}{JM^2} \sum_{j=1}^J \bm{I}^{(j)} }
\end{align}

The observation of phase problem is invariant to the global phase shift $\bm{O} \rightsquigarrow e^{\sqrt{-1} \theta} \bm{O}$.
Consequently, the discrepancy between the true vector $\bm{O}$ and the output $\bm{O}_{\text{alg}}$ from the reconstruction algorithms should be quantified by
\begin{align}
  \text{NMSE} = \frac{1}{N^2} \min_{\theta \in [0, 2\pi)} \| \bm{O} - e^{\sqrt{-1} \theta} \bm{O}_{\text{alg}}\|_2^2
\end{align}
To assess the reconstruction accuracy, we extract the $256 \times 256$ pixel region in the center of the object image, thereby excluding areas not illuminated by the probe.

In blind ptychographic phase retrieval, the solution has additional ambiguities due, for example, to invariance to scaler multiplication $(\bm{O}, \bm{P}) \rightsquigarrow (\beta \bm{O}, \frac{1}{\beta}\bm{P})$ and spatial shifting of images.
For this case, we monitor the convergence of algorithms by simply computing the fitness to the diffraction data, defined as
\begin{align}
  \text{Fitness} = \frac{\sum_{j=1}^J \|\sqrt{\bm{I}^{(j)}} - |\mathsf{F}[\bm{P}_{\text{alg}} \odot \bm{O}_{\text{alg}}^{(j)}]| \|^2_2}{\sum_{j=1}^J \bm{I}^{(j)}}
\end{align}
Here, ($\bm{O}_{\text{alg}}, \bm{P}_{\text{alg}}$) is the output from the reconstruction algorithm, and $\bm{O}_{\text{alg}}^{(j)} \triangleq \mathsf{S}^{(j)}\bm{O}_{\text{alg}}$.

\subsection{Known Probe Case}
We simulated the ptychography measurement of objects $\bm{O}_1$ and $\bm{O}_2$, with probe $\bm{P}_1$ and noise level $\text{SNR} = 30 \ \text{dB}$.
The scan positions are arranged in the Fermat spiral geometry~\cite{FermatSpiral}, and the overlap between scans are changed to control the sampling ratio $\alpha$.
For instance, in Fig. 8, 129 scans of illumination covers 220064 pixels of object image, while collecting $129 \times 128^2$ pixels of diffraction data, corresponding to the sampling ratio of $\alpha \simeq 2.4$.

\begin{figure}[htb]
  \begin{minipage}[b]{1.0\linewidth}
    \centering
    \centerline{\includegraphics[width=4.5cm]{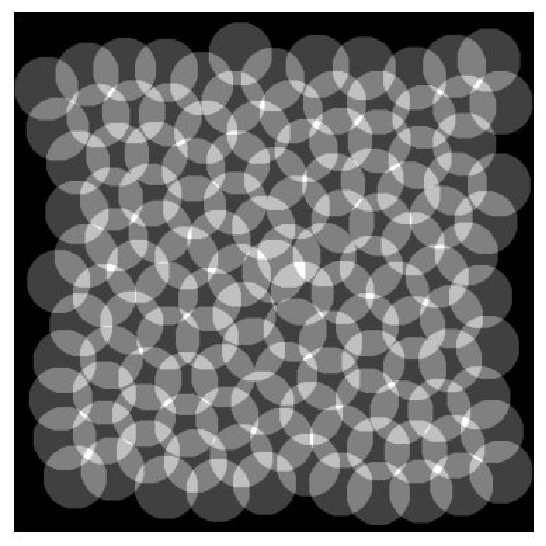}}
    \caption{129 scanning positions arranged in a Fermat Spiral geometry. The number of pixels that is covered by the illumination more than once is 220064.}\medskip
  \end{minipage}
\end{figure}

For the reconstruction of $\bm{O}_1$, Ptycho-EP employs the Gaussian prior $p_{\text{in}}(\bm{O}) = \mathcal{CN}(\bm{O}; \bm{0}, I_N)$, whereas both Gaussian and BG priors are utilized for object $\bm{O}_2$.
NMSE (dB) is presented in Table I, alongside the results obtained for PIE.
The reconstruction error is represented by the the median of NMSE over 10 independent trials, each corresponding to a distinct initial estimate and noise realization.
All of the algorithms are run 400 iterations.

Ptycho-EP not only achieves a lower NMSE compared to PIE but also demonstrates enhanced robustness against small sampling ratio even when approaching the information-theoretic limit.
As visually depicted in Fig. 9, Ptycho-EP with a Gaussian prior yields satisfactory reconstructions for $\alpha \geq 2.3$, which is close to the algorithmic threshold for random orthogonal matrices.
The incorporation of a sparse prior not only improves reconstruction fidelity but also reduces the minimum sampling ratio required for successful phase retrieval.
This strategy holds potential for minimizing the number of measurements in ptychography, thereby decreasing data acquisition time and improving overall efficiency.

\begin{table}[h]
  \centering
  \caption{Reconstruction Error}
  \label{tab:error_comparison}
  \begin{tabular}{c|cc|ccc}
      \toprule
      \multirow{2}{*}{$\alpha$} & \multicolumn{2}{c|}{$\bm{O}_1$} & \multicolumn{3}{c}{$\bm{O}_2$} \\
      \cline{2-6}
      & PIE & Ptycho-EP & PIE & Ptycho-EP& Ptycho-EP \\
      &  & (Gaussian) &  & (Gaussian) & (Sparse)\\
      \midrule
      4.0  & 26.1 & 27.1 & 26.0 & 28.8& 33.4\\
      3.0  & 23.1 & 24.2 & 0.6 & 25.9& 31.6\\
      2.5  & 14.9  & 21.2 & - & 21.8 &30.3\\
      2.4  & 13.4 & 20.2 & - & 20.5& 29.9\\
      2.3  & 11.0 & 19.7 & - & 16.3 &29.7\\
      2.2  & 10.0 & 17.1  & - &3.0& 21.8 \\
      2.1  & 7.9 &13.5 & - & 1.4& 8.5\\
      \bottomrule
  \end{tabular}
\end{table}

\begin{figure*}[t]
  \begin{minipage}[b]{1.0\linewidth}
    \centering
    \centerline{\includegraphics[width=16cm]{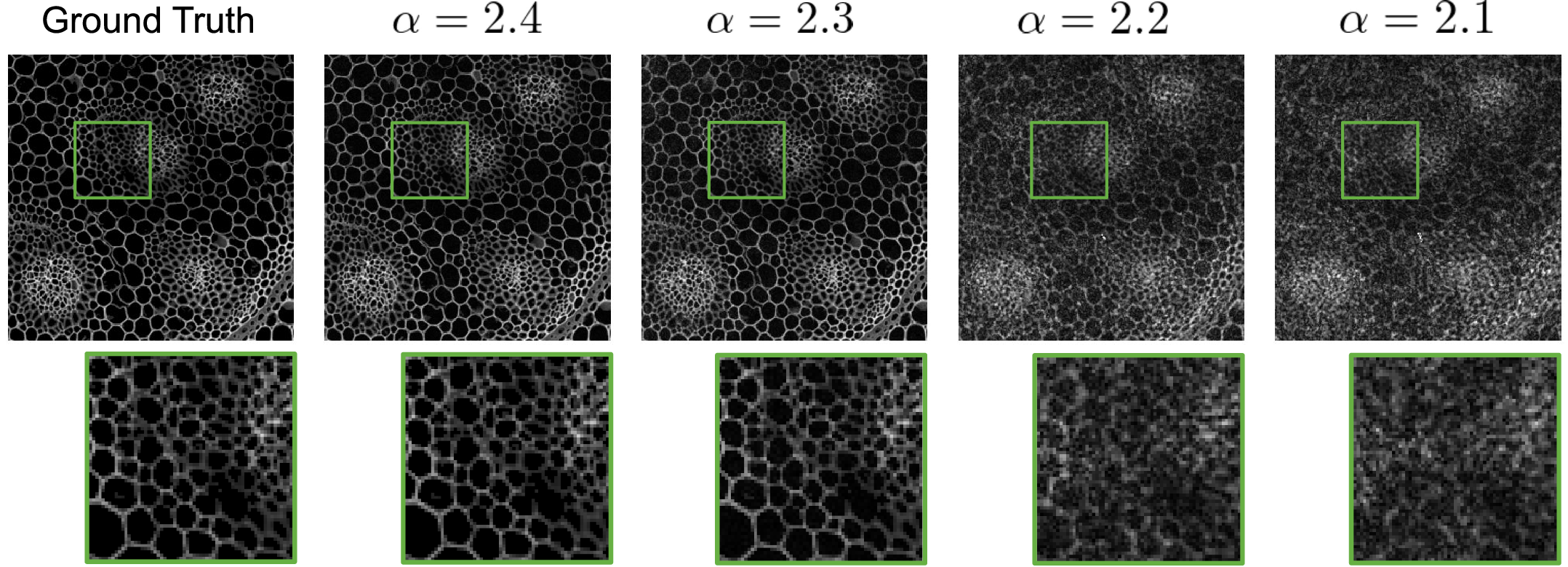}}
    \caption{Reconstructed intensity of $\bm{O}_2$ by Ptycho-EP with Gaussian prior. The sampling ratio $\alpha$ is changed by varying the overlap of scanning positions.}\medskip
  \end{minipage}
\end{figure*}

\begin{figure}[htb]
  \begin{minipage}[b]{1.0\linewidth}
    \centering
    \centerline{\includegraphics[width = 8.2cm]{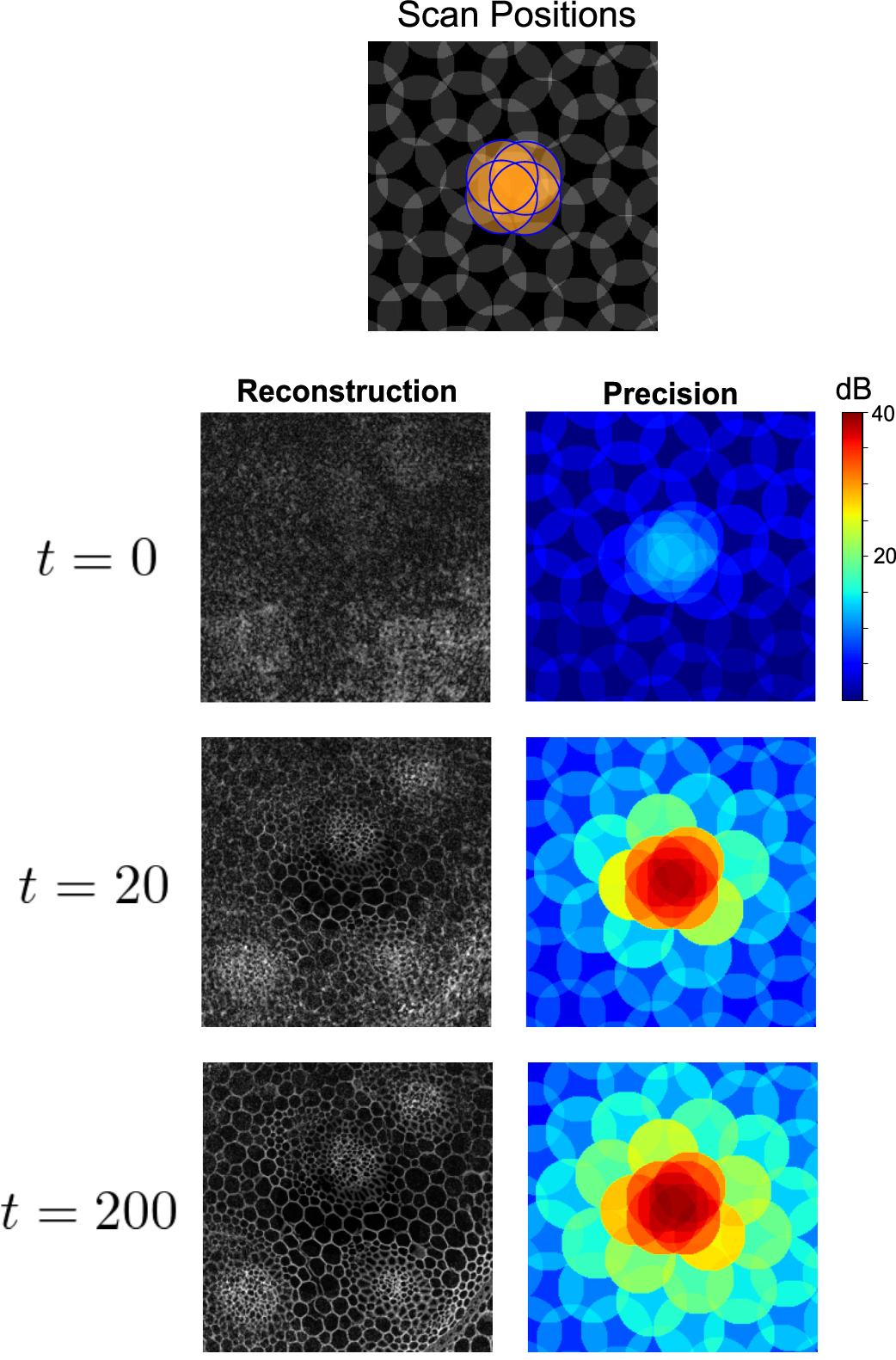}}
    \caption{(Upper) Scanning geometry with non-uniform overlap, generated by adding four scans (orange) to the uniform scanning geometry of Fermat Spiral. (Lower) Reconstruction by Ptycho-EP at iteration number $t = 0, 20, 200$. The left and right image show the evolution of $\widehat{\bm{O}}_{\text{int}}$ and $\widehat{\bm{\Gamma}}_{O, \text{int}}$.}\medskip
  \end{minipage}
\end{figure}

A unique feature of Ptycho-EP is its ability to quantify the uncertainty in reconstruction via the approximate precision $\widehat{\bm{\Gamma}}_{O, \text{int}}$, which provides a pixel-wise estimation of the reliability of the reconstructed image.
As a straightforward demonstration of this method, we illustrate the effect of non-uniform overlap in ptychography, as shown in Fig. 10.
In this simulation, four additional diffraction measurement scans supplement the scans arranged in a Fermat spiral geometry, corresponding to a sampling ratio of $\alpha = 2.2$.
By making large overlap among these four scanning positions, the central region of the object is reconstructed at an early stage of the iterative procedure.
Notably, the reconstructed region expands outward from the center, eventually covering a broader area of the object by iteration $t = 200$.
The evolution of $\widehat{\bm{\Gamma}}_{O, \text{int}}$ reveals that the regions with high reconstruction accuracy grow from the center, aligning with the object reconstruction in the left panel.
In the belief propagation literatures, the role of overlap has been extensively investigated, under the term "spatial coupling" in LDPC codes \cite{SpCp_LDPC,SpCp_LDPC_generalization} and as the "seeding matrix" in compressive sensing \cite{SpCp_CS,SpCp_CS_PRX}.
From this perspective, ptychography can be interpreted as a seeding matrix approach for phase retrieval, where the reconstruction initiates from the "seed"---the region with dense overlaps---and propagates through the overlapping scans.

\subsection{Unknown Probe Case}

\begin{figure*}[t]
  \begin{minipage}[b]{1.0\linewidth}
    \centering
    \centerline{\includegraphics[width=17cm]{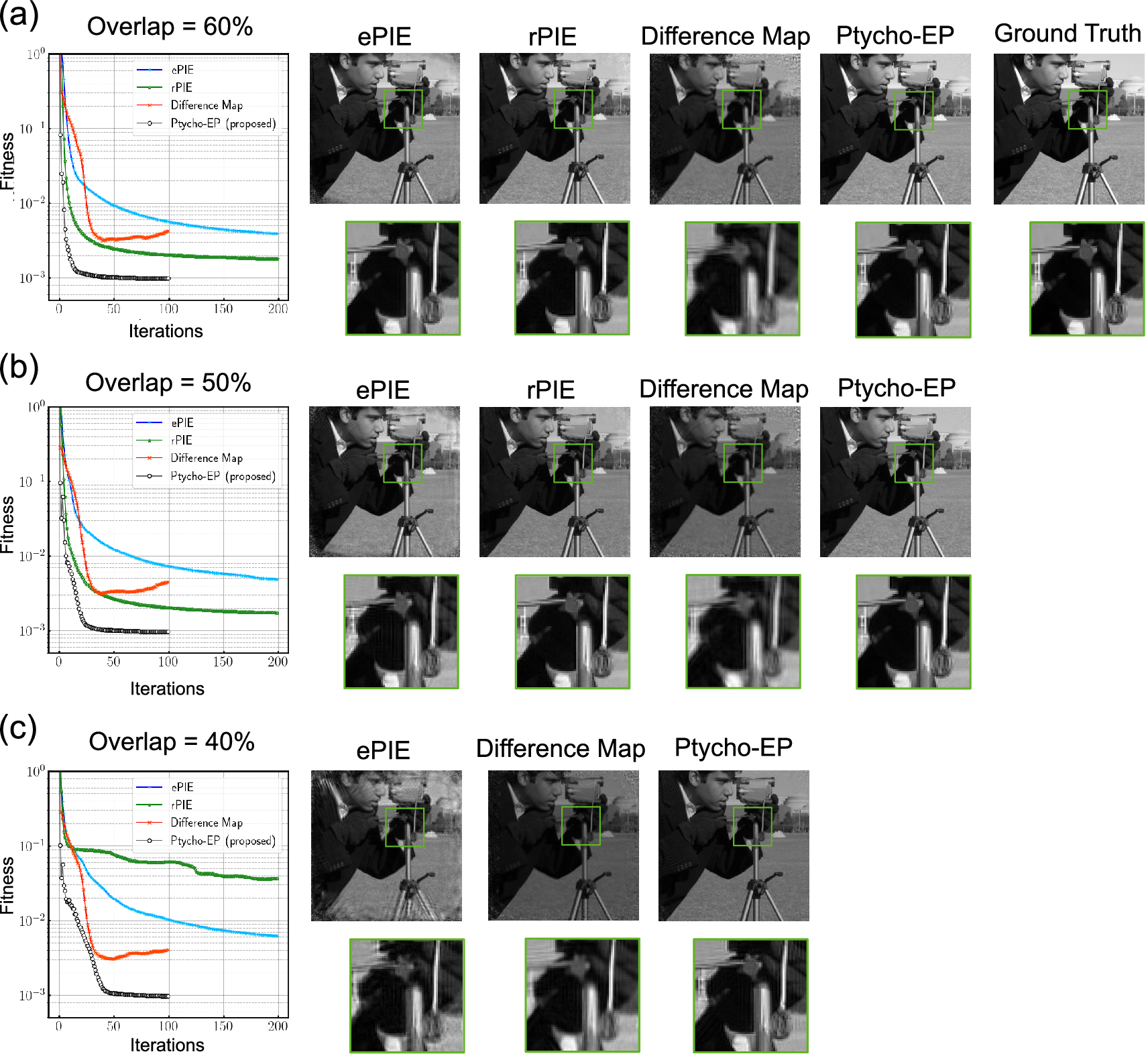}}
    \caption{Reconstruction results of object $\bm{O}_1$ from ptychographic measurement with overlap (a) 60\%, (b) 50 \%, and (c) 40 \%. The convergence behavior is shown as the reduction of the fitness value for four algorithms: ePIE, rPIE, Difference Map, and Ptycho-EP. For each condition, representative reconstructions are displayed on the right. The fitness is represented by the median of 10 independent trials, each corresponding to random initialization by $\mathcal{CN}(\bm{0}, \mathsf{I}_N)$. The rPIE reconstruction for $R = 40 \%$ was unstable, and the typical rPIE reconstruction is omitted in (c).}\medskip
  \end{minipage}
\end{figure*}

We evaluate the performance of Ptycho-EP in the setting of blind ptychographic phase retrieval.
The ground truth for the object and probe is $\bm{O}_1$ and $\bm{P}_2$ in Fig. 7, respectively, while $\bm{P}_2^{\text{init}}$ is used as the initial estimate of the probe.
The scanning geometry is a raster grid pattern with random positional deviations of up to 2 pixels.
We define $d$ as the diameter of the region in the probe image where the intensity exceeds 10\% of the maximum value.
The overlap ratio, denoted as $R$, is given by $R = 1 - \frac{a}{d}$, where $a$ corresponds to the step size of the raster grid.
While the common rule of thumb is that an overlap ratio exceeding 60\% ($R > 60\%$) ensures successful reconstruction~\cite{BUNK2008481}, the design of phase retrieval algorithms that remain robust under conditions of limited overlap presents a practically relevant challenge.

\begin{table}[h]
  \centering
  \caption{Overlap conditions}
  \begin{tabular}{ccc}
      \toprule
       Overlap (\%) & Step Size (pixels) & Number of Scans \\
      \midrule
       60 & 12 & $16 \times 16$ \\
       50 & 15 & $13 \times 13$ \\
      40 & 18 & $11 \times 11$ \\
      \bottomrule
  \end{tabular}
  \label{tab:experiment_conditions}
\end{table}

As shown in Table II, we generated diffraction patterns (SNR = 30 dB) from raster grid scanning using three distinct step sizes.
These correspond to overlap ratios of 60\%, 50\%, and 40\%, with the number of scan positions selected so that the scanning patterns cover the $180 \times 180$ pixel region at the center of the object image.
Ptycho-EP (Gaussian prior) with EM algorithm and Difference Map algorithm are run 100 iterations, since further iterations made no significant progress, whereas rPIR and ePIE are run 200 iterations.
For data with an overlap ratio of $R = 40\%$, the typical computation times were 61 seconds for the Difference Map algorithm, 61 seconds for ePIE, 58 seconds for rPIR, and 47 seconds for Ptycho-EP.

Figure 11 illustrates the convergence behavior of the algorithms alongside typical reconstructed images.
While ePIE and the Difference Map maintain stability across all overlap conditions, their fitness values do not reach the noise level.
With sufficient overlap, rPIE demonstrates accelerated convergence and enhanced denoising performance compared to ePIE and the Difference Map, but becomes unstable when $R=40\%$.
Ptycho-EP exhibits dramatically rapid convergence, reducing the fitness to the noise level within 50 iterations across all overlap conditions.
Although ptychographic phase retrieval algorithms typically require several hundred iterations to achieve a reliable reconstruction, our approach can significantly reduce the computational cost by an order of magnitude.

\section{Conclusions}
We have developed a phase retrieval algorithm for ptychography based on the principles of belief propagation, demonstrating the following points:
\begin{itemize}
  \item Vector Approximate Message Passing, whose convergence relies on the random measurement model, is practically useful for ptychographic reconstruction. The key technique in accelerating and stabilizing the reconstruction is the Stochastic VAMP, derived through belief propagation in a tree graphical model.
  \item When the illumination function of probe is known a priori, ptychographic reconstruction is possible under the sampling ratio approaching the information theoretical limit. Even when the exact probe is unknown, prove retrieval by EM algorithm is possible, providing fast and accurate reconstruction compared to conventional methods.
\end{itemize}

There are modern additions to ptychographic reconstruction algorithms --- for instance, multi-slice modelling of 3D object~\cite{3PIE}, position error correction~\cite{PositionCorrection2012,PositionCorrection2013}, and mixed-states reconstruction~\cite{MixedStates2013}.
Generalization of Ptycho-EP to those problem settings is left for future research.
Parallel implementation of Ptycho-EP, effective use of non-uniform scanning geometry, and exploitation of prior knowledge about the probe are also important directions of further research.

\bibliographystyle{IEEEtran}

\bibliography{refs}

\appendices
\section{Denoisers}
For separable priors (i.e. when $p_{\text{in}}(\bm{O}) = \prod_i p_{\text{in}}(O_i)$), the denoising function is expressed as $\bm{g}_{\text{in}}(\widetilde{\bm{O}}, \widetilde{\bm{\Gamma}}) = (g_{\text{in}}(\widetilde{{O}}_1, \widetilde{{\Gamma}}_1),\cdots, g_{\text{in}}(\widetilde{{O}}_N, \widetilde{{\Gamma}}_N))$.
The scaler denoising function $g_{\text{in}}(\widetilde{O}, \widetilde{\Gamma})$ for BG prior $p_{\text{in}}(O) = \rho \ \mathcal{CN}(O; 0,1) + (1-\rho) \delta(O)$ is given by
\begin{align}
  g_{\text{in}}(\widetilde{O}, \widetilde{\Gamma}) = \pi \times \frac{\widetilde{O}}{1+\widetilde{\Gamma}}
\end{align}
wherein $\pi \in (0,1)$ is the "probability mass of non-zero $O$", definded as
\begin{align}
  \pi = \frac{\rho \ \mathcal{CN}(\widetilde{O}; 0, 1+\widetilde{\Gamma}^{-1})}{\rho \ \mathcal{CN}(\widetilde{O}; 0, 1+\widetilde{\Gamma}^{-1}) + (1 - \rho) \mathcal{CN}(\widetilde{O}; 0, \widetilde{\Gamma}^{-1})} \nonumber
\end{align}
Similarly, the scaler function $g'_{\text{in}}(\widetilde{O}, \widetilde{\Gamma})$ is given as
\begin{align}
  g'_{\text{in}}(\widetilde{O}, \widetilde{\Gamma}) = \pi (1 - \pi) \left|\frac{\widetilde{O}}{1+\widetilde{\Gamma}}\right|^2 + \frac{\pi}{1+\widetilde{\Gamma}}
\end{align}

The denoiser corresponding to the separable likelihood Eq. (6) is derived in~\cite{deepEC} via Laplace approximation.
In our notation, the denosing function in Eq. (31) is expressed as $\bm{g}_{\text{out}}(\widetilde{\bm{\Psi}}, \widetilde{\Gamma}; \bm{I}) = (g_{\text{out}}(\widetilde{\Psi}_1, \widetilde{\Gamma}; I_1),\cdots,g_{\text{out}}(\widetilde{\Psi}_M, \widetilde{\Gamma}; I_M))$, wherein the scaler function $g_{\text{out}}$ is given as
\begin{align}
  g_{\text{out}}(\widetilde{\Psi}, \widetilde{\Gamma}; I) = \frac{\sqrt{I} + 2\sigma^2 \widetilde{\Gamma} |\widetilde{\Psi}|}{1 + 2\sigma^2 \widetilde{\Gamma}} e^{\sqrt{-1} \angle \widetilde{\Psi}}
\end{align}
Likewise, the scaler function $ g'_{\text{out}}(\widetilde{\Psi}, \widetilde{\Gamma}; I)$ is given as
\begin{align}
  g'_{\text{out}}(\widetilde{\Psi}, \widetilde{\Gamma}; I) = \frac{\sqrt{I} + 4\sigma^2 \widetilde{\Gamma} |\widetilde{\Psi}|}{2\widetilde{\Gamma} |\widetilde{\Psi}| (1 + 2\sigma^2 \widetilde{\Gamma}) }
\end{align}

\section{Adaptive EM module}
The adaptive EM approach consists in viewing $\widetilde{\bm{\Psi}}^{(j)}_{\text{int}}\ (j = 1,\cdots,J)$ as an observation of $\mathsf{F}[\bm{P} \odot \bm{O}_{\text{int}}^{(j)}]$ corrupted by the white Gaussian noise with precision $\widetilde{\Gamma}^{(j)}_{\Psi, \text{int}}$.
The log-likelihood of the linear observation is $\sum_j \log \mathcal{CN}(\widetilde{\bm{\Psi}}^{(j)}_{\text{int}} ; \mathsf{F}[\bm{P} \odot \bm{O}_{\text{int}}^{(j)}], \widetilde{\Gamma}^{(j),-1}_{\Psi, \text{int}} I_M)$, and the object of maximization in EM is the expected log likelihood given as
\begin{equation}
  \begin{split}
    &\mathbb{E}\left(\sum_j \log \mathcal{CN}(\widetilde{\bm{\Psi}}^{(j)}_{\text{int}} ; \mathsf{F}[\bm{P} \odot \bm{O}_{\text{int}}^{(j)}], \widetilde{\Gamma}^{(j) -1}_{\Psi, \text{int}} I_M) \right)\\
    &= M \sum_{j=1}^J \log \widetilde{\Gamma}^{(j)}_{\Psi, \text{int}} - \sum_{j=1}^J \widetilde{\Gamma}^{(j)}_{\Psi, \text{int}} \left\| \widetilde{\bm{\Psi}}^{(j)}_{\text{int}} - \mathsf{F}[\bm{P} \odot \widehat{\bm{O}}_{\text{int}}^{(j)}] \right\|_2^2 \\
    &-  \sum_{j=1}^J \widetilde{\Gamma}^{(j)}_{\Psi, \text{int}} \text{tr}(|\bm{P}|^2 \odot \widehat{\bm{\Gamma}}^{(j) -1}_{O,\text{int}}) \nonumber
\end{split}
\end{equation}
Here, expectation is over $\bm{O}_{\text{int}} \sim b(\bm{O}_{\text{int}})$.
Maximizing the right hand side w.r.t. $\bm{P}$ and $\widetilde{\Gamma}^{(j)}_{\Psi, \text{int}}$ yeilds the following update.

\begin{align}
  \bm{P} \leftarrow \frac{\sum_{j=1}^J \widetilde{\Gamma}^{(j)}_{\Psi, \text{int}} \left(\widehat{\bm{O}}^{(j)*} \odot \mathsf{F}^H \widetilde{\bm{\Psi}}^{(j)}_{\text{int}}\right)}{\sum_{j=1}^J \widetilde{\Gamma}^{(j)}_{\Psi, \text{int}} \left(\widehat{\bm{\Gamma}}^{(j) -1}_{O, \text{int}} + |\widehat{\bm{O}}^{(j)}|^2 \right)}
\end{align}

\begin{align}
  \widetilde{\Gamma}^{(j) -1}_{\Psi, \text{int}} \leftarrow \frac{\| \widetilde{\bm{\Psi}}^{(j)}_{\text{int}} - \mathsf{F}[\bm{P} \odot \widehat{\bm{O}}_{\text{int}}^{(j)}] \|_2^2 + \text{tr}(|\bm{P}|^2 \odot \widehat{\bm{\Gamma}}^{(j) -1}_{O,\text{int}})}{M}
\end{align}

Notice that, when assuming that $\widetilde{\Gamma}^{(1)}_{\Psi, \text{int}} = \cdots = \widetilde{\Gamma}^{(J)}_{\Psi, \text{int}}$ and $\widetilde{\bm{\Psi}}^{(j)-1}_{\text{int}} \simeq \bm{0}$, Eq. (45) becomes the probe update rule of Difference Map algorithm.
In the adaptive EM step, Eqs. (45)-(46) is alternately applied $\tau$ times.

\end{document}